\documentclass[%
 reprint,
showpacs,showkeys,
 amsmath,amssymb,
 aps,
]{revtex4-1}

\usepackage{color} 
\usepackage{graphicx}
\usepackage{dcolumn}
\usepackage{bm}

\newcounter{mnotecount}

\newcommand{\mnotex}[1]
{\protect{\stepcounter{mnotecount}}$^{\mbox{\footnotesize $\bullet$\themnotecount}}$ 
\marginpar{
\raggedright\tiny\em
$\!\!\!\!\!\!\,\bullet$\themnotecount: #1} }

\begin{document}

\preprint{APS/123-QED}

\title{Marginally Stable Circular Orbits\\ in Stationary Axisymmetric Spacetimes}

\author{Shabnam Beheshti}
 \altaffiliation{s.beheshti@qmul.ac.uk}
\author{Edgar Gasper\'{i}n}%
 \email{e.gasperingarcia@qmul.ac.uk}
\affiliation{%
 School of Mathematical Sciences, Queen Mary, University of London\\
  Mile End Road, London E1 4NS, United Kingdom}%
  

\begin{abstract}
We derive necessary and sufficient conditions for the existence of marginally stable circular orbits (MSCOs) of test particles in a stationary axisymmetric (SAS) spacetime which possesses a reflection symmetry with respect to the equatorial plane; photon orbits and marginally bound orbits (MBOs) are also addressed.  Energy and angular momentum are shown to decouple from metric quantities, rendering a purely geometric characterization of circular orbits for this general class of metrics.  The subsequent system is analyzed using resultants, providing an algorithmic approach for finding MSCO conditions.  MSCOs, photon orbits and MBOs are explicitly calculated for concrete examples of physical interest.
\end{abstract}


\maketitle


\section{Introduction}

The study of test bodies and light rays in circular orbits around a
central source plays a key role in understanding and predicting
astrophysical phenomena \cite{Pra15, Hod13, Sussman03, ChoPatMalJos12, Damour00, Abramowicz10, Buonnano08}.  Innermost stable circular orbits (ISCOs) of
test particles orbiting massive objects (e.g., neutron stars, Kerr
black holes) have been used to identify transition regions between
inspiralling and merging phases of compact binary systems
\cite{Damour00, Buonnano08}; late dynamical evolution of such binaries
have been fundamental in identifying possible sources of gravitational
waves \cite{Hanna09}.  Stable circular orbits and marginally bound circular orbits play an important role in modelling matter configurations in accretion around axially
symmetric gravitational sources.  The inner radius of the accretion disk
is typically assumed to be equal or close to the ISCO, depending on the luminosity of the
radiating body \cite{Novikov73, Abramowicz10}; in thick-disk models, the cusp of the Roche lobe lies  between the marginally bound circular orbit and a marginally stable circular orbit \cite{Treves88}.


In contrast to Newtonian gravity, in which a spherically symmetric
body admits stable circular orbits for a test mass at any radius, 
predicted behaviour of test masses in General Relativity are highly
varied.  Analysis of timelike and null geodesics in the Schwarzschild
geometry yields a lower bound for the orbital radius of a stable
circular orbit (an ISCO), as well as information on gravitational
redshift and prediction of planetary precession, for instance
\cite{Wal84}; on the other hand the Schwarzschild-de Sitter spacetime additionally admits
an outermost stable circular orbit (OSCO) for test bodies \cite{Kott18}.  When they
exist, OSCOs provide important physical information, namely identifying natural
boundaries between stable and unstable orbits.  Regions between
innermost and outermost stable circular orbits are generally referred
to as marginally stable circular orbits, or MSCOs.  Similarly, particles following parabolic trajectories in black hole spacetimes are typically modeled using marginally bound orbits, or MBOs \cite{Teukolsky83}.

To address the problem of finding such geodesics,
the usual approach relies on a two-Killing field reduction of the spacetime under consideration; this simplifies the problem to the study of one-dimensional particle motion in an effective potential.  In the strong field regime of axially symmetric spacetimes, significant progress has been made in understanding orbits using the well-known effective potential method.  Regions of orbital stability near the horizon of Reissner-Nordstr\"om and in the ergoregion of Kerr have
been used to distinguish black holes from naked singularities
\cite{PugQueRuf11, PugQue15}.  It has also been shown that the
behaviour of MSCOs in these spacetimes differ nontrivially from their
extremal counterparts \cite{PradMaj11, PradMaj13}. 
However, one difficulty of the effective potential method lies in
the fact that the potential is a highly nonlinear function of the
metric components and constants of motion (e.g., energy, angular
momentum).  Consequently, exact calculation of MSCOs quickly becomes
intractable, even in the simplest settings; for example, the MSCO
condition in the Schwarzschild-de Sitter spacetime is a quintic
function of the radius.

To date, the effective potential method has largely been applied on a case-by-case basis.  We exploit the underlying principe of this method, namely that the inner product of a Killing field with a geodesic tangent is constant along the geodesic, to express an MSCO equation for generic stationary, axially symmetric (SAS) spacetimes.  In doing so, we provide a systematic approach for determining \emph{all possible MSCO regions} in a given SAS spacetime for both timelike and null geodesics; analysis of marginally bound orbits (MBOs) follows as a byproduct of our approach.  Consequently, our discussion unifies many of the examples in the literature and provides a simple mechanism by which to study new spacetimes.

Motivation for our current work stems from \cite{OnoSuzFusYamAsa14} and \cite{OnoSuzFusYamAsa15}, in which the authors locate MSCOs for generic static, spherically symmetric spacetimes by using the determinant of an associated linear system.  By generalizing to the stationary, axisymmetric setting, we find that the metric components and test body parameters still decouple in the MSCO equations, just as in static, spherically symmetric case.  However, the associated system is no longer linear, rather it is \emph{algebraic}.  In fact, it is still possible to deduce necessary and sufficient conditions for the existence of an MSCO using standard tools from algebraic geometry.

The paper is organised as follows.  In Section 2, we derive the
geodesic equations which govern the dynamics of both test particle
motion and light rays on the equatorial plane for stationary,
axisymmetric spacetimes.  We restrict our attention to
equatorial circular geodesics since this section captures many of the
primary geometric and dynamic characteristics of the accretion disk,
and is also relevant for Keplerian accretion.  The conditions for existence and stability
of circular orbits are shown to form an algebraic system.  In Section
3, we restrict to the static, axisymmetric case and observe that MSCOs are completely characterized by the norm of their stationary timelike
Killing field.  Static, spherically symmetric examples
are computed, recovering MSCOs for the
Schwarzschild, Reissner-Nordstr\"om and Janis-Newman-Winicour
spacetimes as well as providing a new understanding of the MSCO conditions for the q-metric, appearing in \cite{Gas12Dissertation, BosGasGutQueTok15}.  

In Section 4, we return to the generic stationary, axisymmetric case and introduce a 
generalization of the determinant known as the resultant.  This is a key new element in our analysis.  When paired with physical considerations, a necessary and sufficient condition for the existence of an MSCO is then derived,
namely equation \eqref{necessaryCondition}.  The ISCO for
the Kerr geometry is confirmed, and new examples are calculated including the Pleba\'nski-Demia\'nski family and cylindrically symmetric spacetimes.  In both the static and stationary discussions, we also address photon orbits and MBOs as special cases of our analysis.  We conclude in
Section 5,  
suggesting possible directions for future investigation.  Pseudocode for the devised algorithm is provided in Appendix C.

For the remainder of the paper, we set $G=c=1$.

\section{Stationary axisymmetric spacetimes}
\label{MSCOAxisymmetricSpacetimes}

We start with a stationary, axisymmetric (SAS) spacetime, given in
local coordinates by the Weyl-Lewis-Papapetrou line element
\begin{eqnarray}\label{Weyl-Lewis-Papapetrou}
ds^2 &=& -e^{2U}(dt-\omega d\phi)^2 + e^{-2U}[ e^{2\gamma}(d\rho^2 +
  dz^2) + \rho^2 d\phi^2 ] \nonumber \\
  &=& -A(\rho,z) dt^2 + B(\rho,z) dt d\phi  \nonumber \\
  && \qquad + C(\rho,z) d\phi^2 + D(\rho,z) (d\rho^2 + dz^2). 
\end{eqnarray}
Here we have assumed existence of a pair of commuting Killing fields,
a timelike Killing field $\partial_t$ and a spacelike Killing field
$\partial_\phi$ with $\phi \in [0, 2\pi)$, periodic, so that $U,
  \omega,$ and $\gamma$ are functions of the spatial coordinates $z
  \in\mathbb{R}$ and $\rho \in [0,\infty)$.  For the record, $A(\rho,z) = e^{2U}$, $B(\rho,z)=2\omega e^{2U}$, $C(\rho,z) = \rho^2 e^{-2U} - \omega^2 e^{2U}$, and $D(\rho,z)=e^{2(\gamma-U)}$.

In what follows, we restrict our attention to the $z=z_{0}$ plane; note
that without lost of generality, $z_{0}$ may be set to 0 by redefining the coordinate $z$
appropriately. Furthermore, if the spacetime possess a reflection symmetry
with respect to the $z=0$ plane, then the orbit of a particle with zero transversal
  initial momentum will remain in the plane; the Kerr metric is an example of a spacetime possessing such a symmetry \cite{Wal84}.  Further discussion of this point appears in Appendix A.  We shall refer to the plane of symmetry $z=0$ as the \emph{equatorial plane}.
  
With this assumption, the orbit of a test body in the equatorial plane can be expressed as 
\begin{equation}
d\tau^2 = -e^{2U}(dt-\omega d\phi)^2 + e^{-2U}[ e^{2\gamma}d\rho^2 +
  \rho^2 d\phi^2 ],
\end{equation}
where $\tau$ denotes proper time along a timelike geodesic.  Then, the
Lagrangian of the particle is given by
\begin{equation} \label{lagrangian}
 \mathcal{L} = - A(\rho) \dot{t}^2 + B(\rho) \dot{t}\dot{\phi}
 + C(\rho) \dot{\phi}^2 + D(\rho) \dot{\rho}^2.
\end{equation}
Here the dot refers to differentiation with respect to proper time
$\tau$, and we use the simplified notation $A(\rho) = A(\rho, 0)$ for each of the metric coefficients on the equatorial plane.

Since the spacetime is assumed to be stationary and axisymmetric,  two
constants of motion associated with the specific energy and specific angular momentum of test body are determined by
\begin{eqnarray}
E &=& \frac{1}{2} \frac{\partial\mathcal{L}}{\partial \dot{t}} = -A(\rho)
\dot{t} + \frac{1}{2} B(\rho) \dot{\phi}, \label{eq:t-energy}\\ 
L&=& \frac{1}{2} \frac{\partial \mathcal{L}}{\partial \dot{\phi}} = \frac{1}{2}
B(\rho) \dot{t} + C(\rho) \dot{\phi}.  \label{eq:phi-energy}
\end{eqnarray}
The derivatives $\dot{t}$ and $\dot{\phi}$ are immediately written (noting that $-AC-\frac{1}{4}B^2 = -\rho^2$)
\begin{eqnarray}
\dot{t} &=& \frac{1}{\rho^2} \left(- C(\rho) E +\frac{1}{2}
B(\rho) L \right), \label{eq:tdot} \\ \dot{\phi} &=&
\frac{1}{\rho^2} \left( \frac{1}{2} B(\rho) E + A(\rho) L
\right). \label{eq:phidot}
\end{eqnarray}

Recall that the tangent vectors $v^{a} \equiv {dx^{a}}/{ds}$ to a
geodesic $x^{a}(s)$, have constant norm; we shall write
$g(v,v)=\epsilon$ where $\epsilon=-1$ for timelike geodesics (particles)
parametrised with respect to the proper time $\tau$ and $\epsilon=0$ (or $+1$)
for null (or spacelike) geodesics parametrised with respect to some affine parameter $s$.
Using equations \eqref{eq:tdot} and \eqref{eq:phidot} in \eqref{lagrangian}, we obtain an
equation for the orbit of the test particle:
\begin{eqnarray} \label{eq:Phi-D}
\dot{\rho}^2 &=& \frac{1}{D(\rho)}\left( \epsilon + \frac{1}{\rho^2} C(\rho)E^2 -\frac{1}{\rho^2} B(\rho) EL - \frac{1}{\rho^2} A(\rho) L^2 \right) \nonumber \\
&=& \frac{\Phi(\rho)}{D(\rho)} := -V(\rho).
\end{eqnarray}


We discuss briefly the conditions for a circular orbit in this setup.
Observe that the $\rho$ component of the
geodesic equations for the test particle is given by
\begin{equation}
\ddot{\rho} + \Gamma^\rho_{\alpha\beta}u^\alpha u^\beta = 0,
\end{equation}
where $u^\alpha$ denotes the four-velocity of the particle.  This
equation represents the radial acceleration of the test particle and
may also be found by differentiating $\Phi(\rho)$ in \eqref{eq:Phi-D} with
respect to $\tau$ (assuming $\dot{\rho}\neq 0$)
\begin{equation}\label{eq:potentialderiv}
\ddot{\rho} = -\frac{1}{2} \frac{dV(\rho)}{d\rho}.
\end{equation}

Thus, there are two conditions needed in order to have a circular
orbit.  The first condition is a vanishing radial velocity
$\dot{\rho}(\tau_0)=0$ for some $\tau_0$, giving an initially circular
state of motion (also called a momentarily circular condition); the
second condition is a vanishing radial acceleration
$\ddot{\rho}(\tau_0)=0$, to preserve the circular state of motion
(permanently circular condition).  Setting $\rho_0=\rho(\tau_0)$, the
second condition is equivalent to $dV(\rho_0)/d\rho=0$, by virtue of
\eqref{eq:potentialderiv}.

To investigate linear stability of a circular orbit $\rho_C$, we
consider a perturbation of the form 
$\rho(\tau) = \rho_C(\tau) + \delta(\tau)$.  
Using \eqref{eq:potentialderiv} and the fact that $\dot{\rho}_C=0$,
$\frac{dV(\rho_C)}{d\rho}=0$ and $\ddot{\rho}_C=0$ at $\tau_0$, it is
straightforward to calculate $\ddot{\delta} =
-\frac{1}{2}\frac{dV(\rho)}{d\rho} - \ddot{\rho}_C =
-\frac{1}{2}\frac{d}{d\rho}\left( V(\rho) - V(\rho_C) \right).$ A
Taylor expansion of $V(\rho)$ gives the following relation, to first
order (and near $\tau_0$)
\begin{equation}
\ddot{\delta} = -\frac{1}{2}\frac{d^2V(\rho_C)}{d\rho^2} \delta.
\end{equation}
MSCO conditions for SAS spacetimes identify regions in which stable and
unstable circular orbits reside, represented by
$\frac{d^2V(\rho_C)}{d\rho^2}>0$ and $\frac{d^2V(\rho_C)}{d\rho^2}
<0$, respectively.

Consequently, the set conditions encoding a
marginally stable circular orbit for a test particle are
$\Phi(\rho_0) = 0$, $\Phi'(\rho_0) = 0$, and $\Phi''(\rho_0) = 0$,
where the prime $'$ denotes a derivative respect to $\rho$.  Written explicitly,
the three conditions become 
\begin{widetext}
\begin{eqnarray}
 C(\rho) E^2 - B(\rho) EL - A(\rho) L^2 + \epsilon \rho^2 &=&
 0,  \qquad \label{msco1} \\
 -\left[ \rho C^\prime(\rho) - 2C(\rho) \right] E^2 + \left[ \rho
   B^\prime(\rho) - 2B(\rho) \right] EL + \left[ \rho A^\prime(\rho) -
   2A(\rho) \right] L^2 &=& 0, \label{msco2} \\
-\left[ \rho^2 C^{\prime\prime}(\rho) - 4\rho C^\prime(\rho) +
  6C(\rho) \right] E^2 + \left[ \rho^2 B^{\prime\prime}(\rho) -
  B^\prime(\rho) + 6B(\rho) \right] EL + 
 \left[ \rho^2 A^{\prime\prime}(\rho) - 4\rho A^\prime(\rho) +
   6A(\rho) \right] L^2 &=& 0. \label{msco3}
\end{eqnarray}
We re-express this system of equations in the following compact manner
\begin{equation}\label{eq:fullsystem}
\left[ \begin{array}{cccc} C & -B & -A & \epsilon \rho^2 \\ -\left(
    \rho C^\prime - 2C \right) & \left( \rho B^\prime - 2B \right) &
    \left( \rho A^\prime - 2A \right) & 0 \\ -\left( \rho^2
    C^{\prime\prime} - 4\rho C^\prime + 6C \right) & \left( \rho^2
    B^{\prime\prime} - 4\rho B^\prime + 6B \right) & \left( \rho^2
    A^{\prime\prime} - 4\rho A^\prime + 6A \right) & 0
\end{array} \right]
\left[ \begin{array}{c} E^2 \\ EL \\ L^2 \\ 1 \end{array} \right] =
\left[ \begin{array}{c} 0 \\ 0 \\ 0 \end{array} \right] .
\end{equation}
\end{widetext}
Note that metric coefficient $D(\rho)$ does not appear in \eqref{eq:fullsystem}.

We remark that the conditions for photon orbits and MBOs are encoded in $\Phi(\rho_0) = 0$ and $\Phi'(\rho_0) = 0$.  In the photon orbit setting, we set $\epsilon = 0$ to obtain
\begin{equation}\label{eq:fullphoton}
\left[\! \begin{array}{ccc} C & -B & -A \\
 -\rho C^\prime + 2C & \rho B^\prime - 2B & \rho A^\prime - 2A 
    \end{array} \!\right]\!\!
\left[\! \begin{array}{c} E^2 \\ EL \\ L^2  \end{array} \!\right] \!=\!
\left[\! \begin{array}{c} 0 \\ 0 \end{array} \!\right] .
\end{equation}
In the MBO setting, we set $\epsilon = -1$ and $E=1$ (see, e.g., \cite{Hod13, BarPreTeu72} ) to obtain
\begin{equation}\label{eq:fullMBO}
\left[\! \begin{array}{cccc} C & -B & -A & -\rho^2 \\
 - \rho C^\prime + 2C & \rho B^\prime - 2B & \rho A^\prime - 2A & 0
    \end{array} \!\right]\!\!
\left[\! \begin{array}{c} 1 \\ L \\ L^2 \\ 1 \end{array} \!\right] \!=\!
\left[\! \begin{array}{c} 0 \\ 0  \end{array} \!\right] .
\end{equation}

In \cite{OnoSuzFusYamAsa14}, the authors perform the above analysis for particle orbits in static, spherically symmetric spacetimes.  Their study results in a linear system in $E^2$ and $L^2$, interpreted geometrically as giving conditions for finding a common intersection of three lines in the
$E^2$-$L^2$ plane.  The system is solved by calculating a determinant.  On the
other hand, the current stationary, axially symmetric setup gives rise to an \emph{algebraic} system
of equations in $E^2, EL$ and $L^2$, interpreted
geometrically as giving conditions for finding a common intersection of
\emph{3 conic sections in the $E$-$L$ plane.}  Notice that lines in the $E^2$-$L^2$ plane correspond to degenerate conics in the $E$-$L$ plane, so the procedure put forward in this article
represents a geometric extension of the linear system studied in \cite{OnoSuzFusYamAsa14}.  It is this fact which allows us to analyse the full system using a generalization of the determinant known as the \emph{resultant}; see Section \ref{sec:StationaryCase}.  

Nonetheless, it is interesting to note that in both cases, the geometric information (metric coefficients) are separated from the physical information (particle parameters $E$ and $L$), so that the matrix in
\eqref{eq:fullsystem} depends solely on the spacetime under consideration.

\section{Static case}\label{sec:StaticCase}

In this section, we restrict our attention to the case of static, axially symmetric spacetimes, i.e., $\omega=0$, in order to write down the simplest MSCO equation in this context and motivate our more general calculations.  We demonstrate the simplicity and utility of our approach by recovering classical examples, as well as characterizing circular and bound orbits for a non-spherically symmetric spacetime of more recent interest.

{\bf Marginally Stable Circular Orbits (MSCOs).} Since $B(\rho)$ is assumed to vanish identically, system 
 \eqref{eq:fullsystem}, reduces to
 \begin{multline}\label{eq:staticsystem}
\!\!\!\!\!\left[\!\! \begin{array}{ccc} C & -A & \epsilon \rho^2 \\ - \rho
    C^\prime + 2C  &  \rho A^\prime - 2A  & 0
    \\ - \rho^2 C^{\prime\prime} + 4\rho C^\prime - 6C  &
    \rho^2 A^{\prime\prime} - 4\rho A^\prime + 6A & 0
\end{array} \!\!\right]\!\!
\left[\!\! \begin{array}{c} E^2 \\ L^2 \\ 1 \end{array} \!\!\right]  \\
=
\left[ \begin{array}{c} 0 \\ 0 \\ 0 \end{array} \right].
\end{multline}
Therefore, in the static case a necessary and sufficient mathematical condition for the existence
of nontrivial solutions to \eqref{eq:staticsystem} is the requirement that the above matrix has vanishing
determinant, namely 
\begin{eqnarray}\label{determinantStaticCase}
\epsilon \rho^3 \left[ \rho^2 (A^{\prime\prime}C^\prime - A^\prime
  C^{\prime\prime}) + 2\rho (AC^{\prime\prime} -A^{\prime\prime}C) \right. && \nonumber \\
 \left. + 2 (A^\prime C - A C^\prime) \right] &=&0.
\end{eqnarray}
Every nontrivial solution of the linear system yields a possible candidate for a radius of the MSCO, if it exists.   In physical terms, necessary and sufficient conditions for existence of MSCOs should be solutions to the determinant equation which further satisfy the open conditions $E^2>0$ and $L^2>0$.  One may check these inequalities by substituting the solutions found from \eqref{determinantStaticCase} into the following expressions for $E^2$ and $L^2$, found by solving the first two equations of \eqref{eq:staticsystem}
\begin{eqnarray}
E^2 &=& \frac{-\epsilon \rho^2 (\rho A^\prime - 2A)}{C(\rho A^\prime - 2A) -A(\rho C^\prime - 2C)} \label{eq:Esquared}\\
L^2 &=& \frac{-\epsilon \rho^2 (\rho C^\prime - 2C)}{C(\rho A^\prime - 2A) -A(\rho C^\prime - 2C)} \label{eq:Lsquared}.
\end{eqnarray}

For $\epsilon \neq 0$, \eqref{determinantStaticCase} can be written in terms of the components of the original spacetime as
\begin{equation}\label{MSCOConditionStatic}
\rho U'' + 4\rho^2 U'^3 -6\rho U'^2 + 3U'=0.
\end{equation}

Observe that \eqref{MSCOConditionStatic} only depends on the
metric function $U=U(\rho, z_0)$.  On the other hand, for spacetimes expressed in Lewis-Papapetrou form \eqref{Weyl-Lewis-Papapetrou}, $A=e^{2U}$.  This means that for static, axisymmetric spacetimes, the MSCO is completely characterized by the norm of the stationary timelike Killing vector $\partial_t$.  In the static, spherically symmetric setting, spacetimes expressed locally as
\begin{equation}\label{SphericallySymmetricForm}
ds^2 = -a(r) dt^2 + b(r) dr^2 + c(r) \left( d\theta^2 + \sin^{2}\theta
d\phi^2 \right),
\end{equation} 
satisfy that the norm of this Killing vector corresponds to the metric function
$a(r)$.  As a consequence, we have immediately that 
\begin{equation}\label{NormTimelikeKillingVector}
U(\rho,z)=\frac{1}{2}\ln a(r(\rho, z)).
\end{equation}
Thus, \eqref{MSCOConditionStatic} may be used to calculate MSCOs for static spacetimes in the axially symmetric \emph{or} spherically symmetric settings.

\medskip


{\bf Photon Orbits and Marginally Bound Orbits (MBOs).}  As a by-product of the analysis of system \eqref{eq:staticsystem}, an equation for the location of the photon orbit is obtained using \eqref{eq:fullphoton}.  When $B=0$, the condition is encoded by the determinant $(AC'-CA')=0$, equivalently
\begin{equation}\label{StaticPhotonOrbit}
1-2\rho U'(\rho)=0.
\end{equation}
The stability of such an orbit (if it exists) can be determined
evaluating $\Phi''(\rho)$ at the photon orbit radius $\rho_{ph}$.

Analogously, the MBO is found using \eqref{eq:fullMBO} (with $B=0$) by solving for $L^2$ in both equations to obtain a condition on the metric coefficients, namely
\begin{equation}\label{StaticMBO}
1-2\rho U'(\rho)-e^{2U(\rho)}(1-\rho U'(\rho))=0.
\end{equation}

\medskip

Let us examine \eqref{MSCOConditionStatic} and \eqref{StaticPhotonOrbit} more closely through several concrete examples.  We shall first recover MSCO conditions for some well-known exact solutions of Einstein's equations and end the section with an example of more recent interest, namely the q-metric.  Derivation of the explicit transformation of coordinates between $(t,r,\theta,\phi)$ and $(t,\rho,z,\phi)$ used in subsequent sections may be found in \cite{Wal84}, for instance.

Unless otherwise stated, the range of the coordinates $(t,r,\theta,\phi)$ 
will be $t \in (-\infty,\infty)$, $r \in (0,\infty)$, $\theta \in [0,\pi]$
$\phi \in [0,2\pi)$ while for the Weyl coordinates $(\rho,z)$, we will
consider  $\rho \in (0,\infty )$ and $z\in (-\infty,\infty)$.  Further discussion on Weyl canonical coordinates may be found in \cite{GriPod09}.

\subsection{Schwarzschild spacetime}

The Schwarzschild spacetime in Weyl-Lewis-Papapetrou form may be
written as
\begin{equation}\label{eq:schw}
ds^2_{Sch} = - \frac{L-M}{L+M} dt^2 + \frac{L+M}{L-M} \rho^2 d\phi^2
+\frac{L+M}{\ell_{+}\ell_{-}} ( d\rho^2 + dz^2) ,
\end{equation}
where $L = \frac{1}{2}(\ell_{+}+\ell_{-})$ and $\ell_{\pm} =
\sqrt{\rho^2 + (z\pm M)^2}$.  Here, $M$ denotes the ADM mass.  In the notation of Section
\ref{MSCOAxisymmetricSpacetimes}, $A(\rho,z) = \frac{L-M}{L+M}$, so
\begin{equation}
U(\rho,z) = \frac{1}{2}\ln \left(\frac{\ell_{-}+z-M}{\ell_{+}+z+M}
\right).
\end{equation}
In these coordinates, the equatorial plane corresponds to $z_0=0$,
whereby the MSCO condition \eqref{MSCOConditionStatic} reduces to
\begin{equation}\label{eq:rhoSchw}
 \rho_{\mbox{\tiny MSCO}}=2\sqrt{6}M.
\end{equation}
Using the change of coordinates 
$r = L+M , \, z = (r-M)\cos\theta , \, \rho = \sqrt{r^2
  -2Mr}\sin\theta $,
the metric is rewritten in standard form
\begin{eqnarray}\label{SchwarzschildStandardForm}
ds^2 &=& - \left( 1-\frac{r_g}{r} \right) dt^2 + \left( 1-\frac{r_g}{r}
\right) ^{-1} dr^2 \nonumber \\
&& \qquad + r^2 \left( d\theta^2 + \sin^2 \theta d\phi^2
\right),
\end{eqnarray}
where $r_g=2M$.  In terms of these spherically symmetric coordinates, \eqref{eq:rhoSchw} yields the well-known
result 
$r_{\mbox{\tiny MSCO}}=6M = 3 r_g$. 

\medskip

Location of the photon orbit in Schwarzschild spacetime may also be calculated easily, as \eqref{StaticPhotonOrbit} reduces to 
\begin{equation}
\rho_{\mbox{\tiny ph}}=\sqrt{3}M,
\end{equation}  
which corresponds to $r_{\mbox{\tiny ph}}=3M$, as expected.  By \eqref{StaticMBO}, the MBO analogously reduces to $\rho_{\mbox{\tiny MB}} = 2\sqrt{2} M$, equivalently $r_{\mbox{\tiny MB}}=4M$, the well-known classical bound orbit.

We note that although derivations of these results using the effective potential method may be found, our approach provides an efficient way to ``read off'' all the orbit conditions using relatively simple operations on the metric coefficients.

\subsection{Reissner-Nordstr\"om spacetime}

The Reissner-Nordstr\"om spacetime in Weyl-Lewis-Papapetrou form may
be written as
\begin{eqnarray}\label{ReissnerNordstroemWeylFrom}
ds^2_{RN}&=& - \frac{(\ell_{+} + \ell_{-})^2-4d^2}{(\ell_{+} + \ell_{-}
  + 2M)^2} dt^2 + \frac{(\ell_{+}+\ell_{-}+2M)^2}{(\ell_{+} +
  \ell_{-})^2 -4d^2}\rho^2 d\phi^2 \nonumber \\
  && \qquad 
+ \frac{(\ell_{+}+\ell_{-}+2M)^2}{4\ell_{+}\ell_{-}}\left(d\rho^2 +
dz^2\right),
\end{eqnarray}
where $\ell^{2}_{\pm} = \rho^2 + (z \pm d)^2$, $d^2 = M^2-e^2$, and
$e$ represents the electric charge of a body having (ADM) mass $M$.
See \cite{GriPod09}, for instance.

Following the notation of Section
\ref{MSCOAxisymmetricSpacetimes}, we identify
\begin{equation}
 U(\rho,z) = \frac{1}{2}\ln \left(\frac{(\ell_{+}+\ell_{-})^2
   -4d^2}{(\ell_{+}+\ell_{-} +2M)^2} \right),
\end{equation}
and evaluate \eqref{MSCOConditionStatic} at the equatorial plane
$z_0=0$.  The MSCO condition becomes
\begin{multline}
4\rho^2 (d^2 +\rho^2)^{-3/2}\left( M + \sqrt{d^2 +\rho^2} \right)^{-3} \Big{\{}-4d^4  \\
+ M\rho^2\left(-3M + \sqrt{d^2 + \rho^2} \right. \\
\left. -4d^2M(M+ \sqrt{d^2 + \rho^2}) \right)\Big{\}} =0.
\end{multline}
Observe that $\rho \neq 0$ and the denominator never vanishes, so substituting $d=M^2 - e^2$ reduces the above to
\begin{multline}\label{MSCORNWeylCoord}
 -3M^2\rho^2 -8M^3\sqrt{M^2-e^2+\rho^2}
\\ +8Mq^2 \sqrt{M^2 -e^2 + \rho^2}+ M\rho^2 \sqrt{M^2-e^2 +\rho^2}\\
-8M^4 + 12M^2e^2 -4e^4=0.
\end{multline}
Setting $x^2= M^2-e^2+\rho^2$, \eqref{MSCORNWeylCoord} may be rewritten as a
cubic equation in $x$ 
\begin{equation}
Mx^3 -3M^2x^2-9M(M^2-e^2)x+(5M^2-e^2)(M^2-e^2) = 0,
\end{equation}
the roots of which may be explicitly found using a variety of
techniques.
%

We remark that \eqref{MSCORNWeylCoord} compares directly with known MSCO
conditions.  To see this, we use
the change of coordinates $r = M + \frac{1}{2}\left(\ell_{+}+
\ell_{-}\right)$, $\cos \theta = \frac{1}{2d}\left(\ell_{+}+ \ell_{-}
\right)$ to recast the Reissner-Nordstr\"om metric in the typical
static, spherically symmetric form
\begin{eqnarray}\label{ReissnerNordstroemSpherical}
 ds^2&=&-\left(1+ \frac{r_g}{r}+ \frac{e^2}{r^2} \right) dt^2 + \left(1+
 \frac{r_g}{r} + \frac{e^2}{r^2}\right)^{-1} dr^2 \nonumber \\
 && \qquad + r^2 \left(
 d\theta^2 + \sin^2 \theta d\phi^2 \right).
\end{eqnarray}
Here $r_g = 2M$.  Note that $\rho=\sqrt{r^2-2Mr +q^2}$ when $z_0=0$, so condition
\eqref{MSCORNWeylCoord} can be re-expressed in terms of the areal
radius $r$, namely
\begin{equation}\label{MSCORNSpherical}
r_g r^3 - 3r^2_g r^2 + 9 e^2 r_g r - 8 e^4 = 0.
\end{equation}
A particular case which can be readily checked is the extremal
Reissner-Nordstr\"om spacetime ($M=e$). In this case, using
\eqref{MSCORNWeylCoord} renders $ \rho = 3M$ which in term of the
areal radius corresponds to $r=4M$, in agreement with the
results given in \cite{ OnoSuzFusYamAsa14, PugQueRuf11}.

\medskip

The location of the photon orbit encoded in
\eqref{StaticPhotonOrbit} gives
\begin{equation}
\rho_{\mbox{\tiny ph}}=\sqrt{d^2+M^2/2 \pm (M/2)\sqrt{8d^2 +M^2}},
\end{equation}
which corresponds to $r_{\mbox{\tiny ph}}=\frac{1}{2}(3M \pm
\sqrt{9M^2-8e^2})$, as expected from \cite{PugQueRuf11}.  Finally, the MBO condition in \eqref{StaticMBO} becomes $Mr^3-4M^2r^2+4Me^2r-e^4=0$.

\subsection{Janis-Newman-Winicour spacetime}

  
In the case where the spacetime under consideration is naturally
expressed in static, spherically symmetric form (as opposed to the Weyl-Lewis-Papapetrou form \eqref{Weyl-Lewis-Papapetrou}), it is still possible
to use \eqref{MSCOConditionStatic} as an alternate means of
calculating the MSCO condition.  We demonstrate this using the
Janis-Newman-Winicour metric, most commonly written in local
coordinates as 
\begin{eqnarray}
ds^2_{JNW} &=& -\left( 1-\frac{2M}{\gamma r}\right)^\gamma dt^2 + \left(
1-\frac{2M}{\gamma r}\right)^{-\gamma} dr^2 \nonumber \\
&& \!\!+ \left( 1-\frac{2M}{\gamma
  r}\right)^{1-\gamma} r^2 \left(d\theta^2 + \sin^2 \theta d\phi^2
\right).
\end{eqnarray}
We make use of the discussion in Section \ref{sec:StaticCase} and the construction
of Weyl coordinates for spherically symmetric spacetimes (see, e.g., \cite{Wal84}).  First,
using the definition of $\rho$, it is straightforward to show that
\begin{equation}
\rho= r\sqrt{1-\frac{2M}{\gamma r}}\sin \theta.
\end{equation}
Restricted to the equatorial plane, this equation may be used to
re-express $r$ as a function $r=r(\rho,z)$, namely
 $r= \frac{M}{\gamma}+ \sqrt{\rho^2 + \left(\frac{M}{\gamma}\right)^2}$.
Using  this form of $r$ and equation \eqref{NormTimelikeKillingVector}, the metric function
$U(\rho,z)$ at the equatorial plane $z=0$ is given by
\begin{equation}\label{UforJNWmetric}
U(\rho) = \frac{\gamma}{2}\ln \left( \frac{-1 + \sqrt{
    (\gamma\rho/M)^2 + 1}}{1+ \sqrt{(\gamma \rho/M)^2 + 1}}\right).
\end{equation}
Direct computation of \eqref{MSCOConditionStatic} gives rise to the
following MSCO equation
\begin{equation}
\frac{4 \gamma \rho^2 \left(\gamma^2 \rho^2 + M^2 \left( 2 + 4
  \gamma^2 -6\gamma \sqrt{1 + \frac{\gamma^2 \rho^2}{M^2}}\right)
  \right)}{(M^2+ \gamma^2
  \rho^2)\sqrt{1+\frac{\gamma^2\rho^2}{M^2}}}=0.
 \end{equation}
Since the denominator of the last expression is never vanishing and
$\rho \neq 0$, the location of the MSCO is found as
\begin{equation}\label{MSCOJNWrho}
\rho_{\tiny \mbox{MSCO}} = M\sqrt{14-\frac{2}{\gamma^2} \pm
  \frac{\sqrt{\gamma^6 (5\gamma^2-1)}}{\gamma^4}}.
\end{equation}
Recasting the final expression in terms of $r$ via 
yields
$r_{\tiny\mbox{MSCO}}=\frac{M}{\gamma}\left(1+ 3\gamma \pm
\sqrt{5\gamma^2-1} \right)$,
 in direct agreement with the result
calculated in \cite{ChoPatMalJos12}.
Additionally, using \eqref{StaticPhotonOrbit} we easily find the
location of the photon orbit to be
\begin{equation}
\rho_{\mbox{\tiny ph}}=\frac{M}{\gamma}\sqrt{4\gamma^2-1}.
\end{equation}
The MBOs may found by solving
\begin{equation}
\frac{1+W}{1-W}(1-W^\gamma)+\gamma W^\gamma-2\gamma=0,
\end{equation}
where $W=\left( 2M^2 +\gamma^2\rho^2 - 2M\sqrt{M^2+\gamma^2\rho^2}\right)/\gamma^2\rho^2$.

\subsection{The q-metric}

Condition \eqref{MSCOConditionStatic} may also be used to directly
determine the marginally stable circular orbits of
\emph{non-spherically symmetric} examples as well.  An insightful example to
consider is the family of static vacuum spacetimes derived in
\cite{Que10} and \cite{Que12} known as the $q$-metric.  
MSCOs obtained in this section may be contrasted with the those found using the effective potential method in \cite{Gas12Dissertation, BosGasGutQueTok15}.  Notably, we find several new candidates for orbits not yet appearing in the literature and remark on their nature.

In Schwarzschildean coordinates 
the line element of the $q$-metric reads
\begin{multline}\label{q-metric}
ds^2_{q}=-\left(1-\frac{2m}{r}\right)^{\!\!1+q}dt^2 \\
+\left( 1-\frac{2m}{r}
\right)^{\!\!-q}\!\Bigg[\!\left(1+ \frac{m^2\sin^2\theta}{r^2-2mr}
  \right)^{\!\!-q(2+q)} \!\!\left(1-\frac{2m}{r} \right)^{\!\!-1} \!\!dr^2 \\ +
  \left(1+ \frac{m^2\sin^2\theta}{r^2-2mr} \right)^{\!\!-q(2+q)}
  r^2d\theta^2 + r^2\sin^2 \theta d\phi^2 \Bigg].
\end{multline}
Here $m$ is a positive parameter and $q>-1$.  This solution represents
the gravitational field of a static mass with quadrupole moment in
empty space; it reduces to the Schwarzschild metric in the limit $q
\rightarrow 0$.  The ADM mass for this spacetime is
  $M_{ADM} = m(1+q)$, the quadrupole moment is given by
  $M_{2}=-m^3q(1+q)(2+q)/3$,  and higher multipole moments are proportional to
  $mq$.  As such, $q$ can be interpreted as a
  parameter measuring deviations from spherical symmetry.  

Originally constructed in \cite{Que10} by applying a Zipoy-Vorhees transformation to the Schwarzschild metric, the \emph{a priori} range of validity of the coordinates appears to be that of Schwarzschild, namely $t \in (-\infty,\infty)$, $r \in (0,\infty)$, $\theta \in [0,\pi]$, $\phi \in [0,2\pi)$.  However, an interesting feature of this family of spacetimes is the appearance of naked singularities.  It is known that despite having a Killing horizon at $r=2m$, for any value $q \neq 0$, $r=2m$ is also a curvature singularity. Moreover, for $q \in  (-1,-1+\sqrt{3/2}]\backslash\{0\}$ there are further curvature singularities located at $r=m(1\pm \cos\theta)$.  These singular regions intersect the equatorial plane $\theta=\pi/2$ at $r=2m$, $r=m$ and $r=0$ \cite{Que10,BosGasGutQueTok15}.

Consequently, after fixing $q$ the range of validity of the coordinates must be reduced to ensure the metric is well defined.  Observe that the norm of the timelike Killing vector is $(1-{2m}/{r})^{1+q}$.  This means that for a given, fixed $q$, we must restrict the radial coordinate $r$ in order to ensure a real quantity for this norm (timelike or spacelike).  For example, if $q=-1/2$, we require $r \in (2m,\infty)$, whereas for $q =-1/3$, we require $r \in (0, \infty)\setminus \{m,2m\}$.  In what follows, we consider a generic radial coordinate $r$, keeping in mind possible restrictions on admissible choices for $q$.

We first express \eqref{q-metric} in Lewis-Papapetrou form by
introducing coordinates 
\begin{eqnarray}
 r&=&m+\frac{1}{2}\big( \sqrt{\rho^2 + (z+m)^2} + \sqrt{\rho^2 +
    (z-m)^2} \big),
\label{rho-qmetric} \nonumber \\ 
\cos \theta &=& \frac{1}{2m} \big( \sqrt{\rho^2 + (z+m)^2} -
\sqrt{\rho^2 + (z-m)^2} \label{z-qmetric} \big). 
\end{eqnarray}

Notice at $z=0$, that $\theta=\pi/2$ (for any $\rho$).  Recasting the relation
between $\rho$ and $r$ in the more familiar form 
\begin{equation}\label{rho-qmetric-familarForm}
\rho^2 = (r^2-2mr) \sin^2 \theta, 
\end{equation}
we use \eqref{rho-qmetric} to obtain
\begin{equation}
U(\rho,z )=\frac{1+q}{2}\ln \left(
\frac{\ell_{+}+\ell_{-}-2m}{\ell_{+}+\ell_{-}+2m}\right),
\end{equation}
where 
$\ell_{\pm}=\rho^2 + (z\pm m)^2$.
Evaluating the condition encoded in \eqref{MSCOConditionStatic} at $z=0$ renders the following equation for the location of the MSCO
\begin{multline}
4m(1+q)(m^2 +\rho^2)^{-3/2}\rho^2\left(m^2(6+4q(2+q)) \right.  \\
\left. + \rho^2 - 6m(1+q)\sqrt{m^2 +\rho^2}\right)=0.
\end{multline}
Since $m>0$ and $q>-1$, the above equation reduces to 
\begin{equation}\label{MSCOqMetricRho}
m^2(6+4q(2+q)) + \rho^2 - 6m(1+q)\sqrt{m^2 +\rho^2}=0.
\end{equation}
In light of \eqref{rho-qmetric-familarForm}, we invoke our prior discussion, noting that there are values of $q$ for which the metric in \eqref{q-metric} remains well-defined for $r$ in some subinterval of $(0, 2m)$ and restore the original coordinates in order to complete the analysis; this is further made possible by the fact that only even powers of $\rho$ appear in the equation.  
Rewriting condition \eqref{MSCOqMetricRho} in terms of $r$ (for $z=0$) we obtain
\begin{equation}\label{MSCOrcomplete}
m^2(6+4q)(2+q)+ r^2 -2m(3(1+q)\sqrt{(m-r)^2} +r)=0.
\end{equation}
Assuming that $r>m$, a simpler equation is obtained, namely
\begin{equation}
2m^2(2+q)(3+2q)-2m(4+3q)r + r^2=0,
\end{equation}
whose solutions are
\begin{equation}\label{SolObtainedInUNAMThesis}
r_{\tiny \mbox{MSCO}} = m(4 + 3q \pm \sqrt{4 +10 q + 5q^2}).
\end{equation}
This last result is consistent with the effective potential analysis made in \cite{BosGasGutQueTok15}
 where the MSCO was found by restricting to the region $r>2m$.  In particular, our result proves that  \emph{the only MSCOs in the region $r>m$ must be the two found in} \eqref{SolObtainedInUNAMThesis}.


Let us return to the discussion of MSCO for a test particle and now assume $r<m$.  In this case \eqref{MSCOrcomplete}
renders the following additional candidates
\begin{equation}\label{NewSolutionqMetric}
\tilde{r}_{\tiny \mbox{MSCO}} = -m(2 + 3q \pm \sqrt{4 +10 q + 5q^2}).
\end{equation}
By choosing the first sign in \eqref{NewSolutionqMetric}, $\tilde{r}_{\tiny{\mbox{MSCO}}}<0$
so this radius is disregarded since we restrict to positive $r$.

Choosing the second sign in \eqref{NewSolutionqMetric} and assuming $q\in(-1/2,0)$, we have $\tilde{r}_{\tiny\mbox{MSCO}}>0$.  However, one expects the photon orbit to impose a lower bound on the location of circular orbits (see \cite{Gas12Dissertation, BosGasGutQueTok15}), making this root unphysical; this can be seen by checking whether the MSCO sufficiency conditions $E^2>0$ and $L^2>0$ are satisfied.  For completeness, we do this now, noting that other examples in this article may be studied in a similar fashion.

Using \eqref{eq:Esquared} and \eqref{eq:Lsquared}, and assuming $r<m$, we compute $E^2$ and $L^2$ to be
\begin{eqnarray}
E^2 &=& \frac{r(m-m(1+q)-r)(1-\frac{2m}{r})^{-q}}{(r-2m)(m-2m(1+q)-r)} \\
L^2 &=& \frac{m(1+q)(r-2m)^2(1-\frac{2m}{r})^q}{m-2m(1+q)-r}.
\end{eqnarray}
Observe that for $0<r<m$, there are infinitely many admissible $q \in (-1/2,0)$ for which these equations may be evaluated, e.g., $q=-1/3$.  However, substituting $\tilde{r}_{\tiny \mbox{MSCO}}$ from \eqref{NewSolutionqMetric} into the first equation gives $E^2<0$, making the candidate unphysical.  Note however, that $L^2>0$ for this root.

Thus, of  the four possible radii/roots of \eqref{MSCOrcomplete}, two are deemed unphysical, leaving only those in \eqref{SolObtainedInUNAMThesis} which satisfy $r \neq (2q+3)m$.  See Figure \ref{fig:qmetricMSCO}.  It is important to note two items.  The method devised in this paper finds \emph{all possible candidates} for MSCOs--an algorithmic advantage over the traditional effective potential method, which did not detect these roots.  The open conditions $E^2>0$, $L^2>0$ provide a simple mechanism by which to determine the physicality (or disregard) each candidate root, not only confirming the orbits found in \cite{Gas12Dissertation, BosGasGutQueTok15}, but closing the possibility for other possible MSCOs.    We consolidate our calculations in Figure \ref{fig:qmetricMSCO}.

%

\begin{figure}[h]
\includegraphics[width=0.4\textwidth]{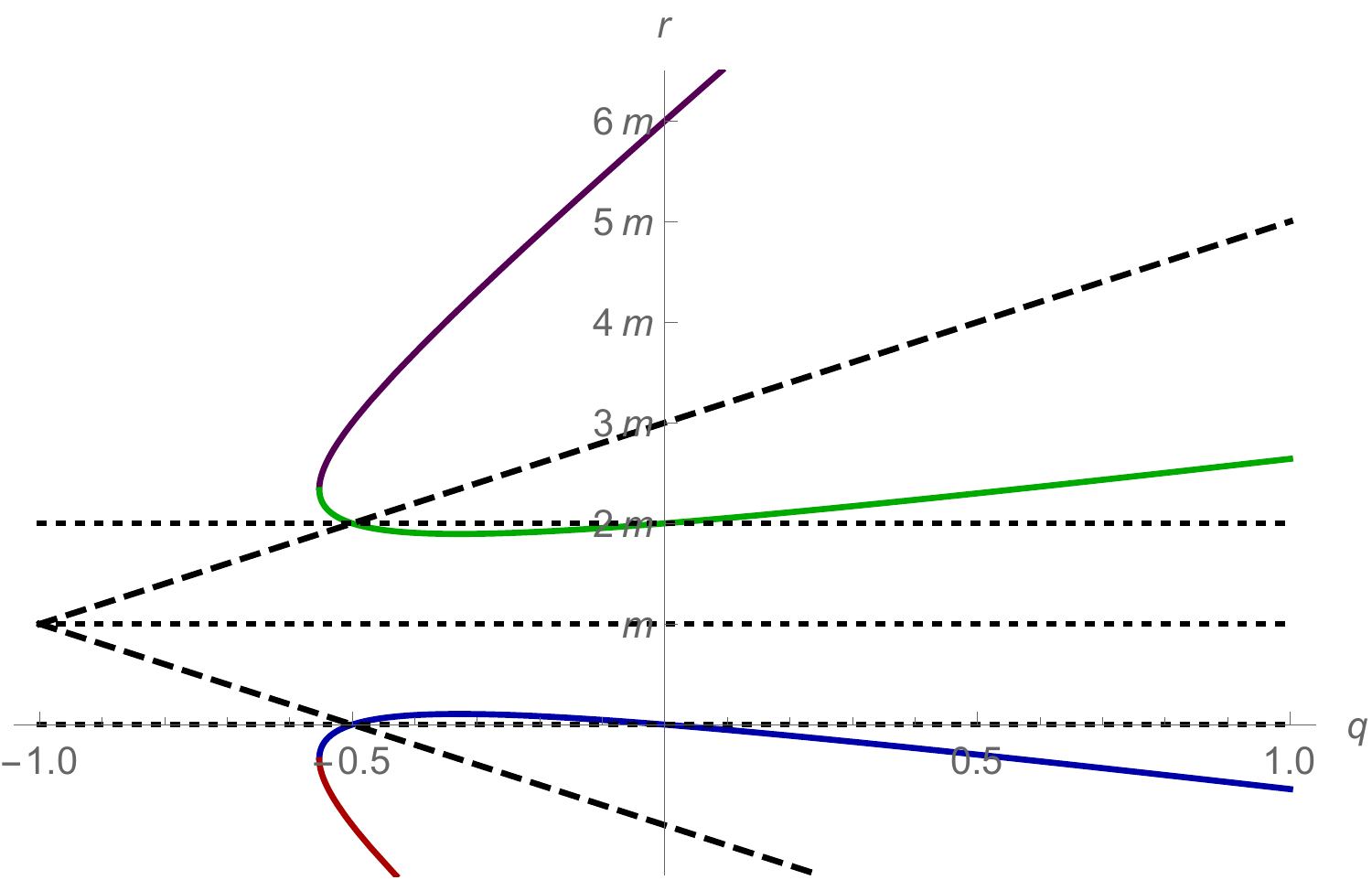}
\caption{\label{fig:qmetricMSCO} MSCO candidates for the $q$-metric in the equatorial plane.
 The two roots for $r_{\tiny\mbox{MSCO}}(q)$ are shown in the upper branch of the graph (red and green), and  
the two roots $\tilde{r}_{\tiny\mbox{MSCO}}(q)$  are plotted in the lower branch of the graph (blue and red).  The dotted horizontal lines represent the locations of the singular regions $r=0$, $r=m$ and $r=2m$.  The dashed line represents the locations $\pm m - 2m(1+q) \mp r=0$ for which $E^2$ and $L^2$ diverge.  Physically relevant MSCO candidates are those above the dashed line}
\label{fig:qmetricMSCO}
\end{figure}

We remark that just as in the previous examples, the location of the photon orbit is found via \eqref{StaticPhotonOrbit} to be
\begin{equation}
\rho_{\mbox{\tiny ph}}=m\sqrt{4q^2 + 8q +3}.
\end{equation}
This corresponds to $r_{\mbox{\tiny ph}}=m(3 + 2q)$, which is also in
agreement with the results given in \cite{BosGasGutQueTok15}; see also \cite{Gas12Dissertation}.  Furthermore, conditions for marginally bound orbits in the $q$-metric spacetime have not previously been calculated in the literature, but take the surprisingly simple form
\begin{equation}
r_{\mbox{\tiny MB}}=\frac{2m(2+q)}{1-q}.
\end{equation}

\section{Stationary case}\label{sec:StationaryCase}

We now return to the general stationary, axially symmetric case, where $\omega \neq 0$ (equivalently $B(\rho,z) \neq 0$) in \eqref{Weyl-Lewis-Papapetrou}.  The equations in \eqref{eq:fullsystem}, determining existence of MSCOs no longer reduce to the \emph{linear} system in \eqref{eq:staticsystem}; fortuitously, the metric components remain decoupled from the particle
parameters, giving rise to an \emph{algebraic} system, namely the algebraic system in $E$ and $L$ appearing in \eqref{msco1}--\eqref{msco3}.  We would like to determine the nature of the solutions to this system, doing so by using a generalization of the determinant known as the resultant.  This will give us a condition for existence of solutions to the algebraic system, on which we may further impose $E^2>0$ and $L^2>0$ in order to obtain necessary and sufficient physical conditions for existence of MSCOs.

\subsection{Resultants}
In this subsection, we define the resultant in two settings and discuss several pertinent facts on
solving systems of polynomials; we restrict our attention to those definitions and results which will be pertinent in our subsequent calculations.  More general theorems on solving polynomial systems may be found in standard algebraic geometry literature; accessible and complete treatment of  our discussion may be found in \cite{Cox} and \cite{Sturmfels}.

{\bf Particle Orbits: polynomials of three variables.}  Let $F(x,y,z)$, $G(x,y,z)$ and $H(x,y,z)$ be homogeneous polynomials in the variables $x$, $y$, and $z$ defined as
\begin{eqnarray}\label{eq:FGH}
F &=& a_0 x^2 + a_1 xy + a_2 y^2 + a_3 xz + a_4 yz + a_5 z^2 , 
\nonumber \\ 
G &=& b_0 x^2 + b_1 xy + b_2 y^2 + b_3 xz + b_4 yz
+ b_5 z^2 ,  \\
H &=& c_0 x^2 + c_1 xy + c_2 y^2 + c_3 xz + c_4 yz
+ c_5 z^2. \nonumber
\end{eqnarray}
The Jacobian determinant of $F,G$ and $H$, given by
\begin{equation}
J= J(x,y,z) = \det \left[ \begin{array}{ccc} F_x & F_y & F_z \\ G_x &
    G_y & G_z \\ H_x & H_y & H_z \end{array} \right],
\end{equation}
is a polynomial of degree 3 in the variables.  Here, subscripts denote
partial derivatives (e.g., $F_x = \frac{\partial F}{\partial x}$).
Consequently, the partial derivatives of $J$ are quadratics denoted by
\begin{eqnarray}
J_x &=& u_0 x^2 + u_1 xy + u_2 y^2 + u_3 xz + u_4 yz + u_5 z^2 , \nonumber \\
 J_y
&=& v_0 x^2 + v_1 xy + v_2 y^2 + v_3 xz + v_4 yz + v_5 z^2 , \\
 J_z &=&
w_0 x^2 + w_1 xy + w_2 y^2 + w_3 xz + w_4 yz + w_5 z^2. \nonumber
\end{eqnarray}
With this notation in place, the resultant of $F,G$ and $H$ is given
by the following determinant
\begin{equation}\label{eq:resultant}
\mbox{Res}(F,G,H) = \det \left[ \begin{array}{cccccc} a_0 & b_0 & c_0
   & u_0 & v_0 & w_0 \\ a_1 & b_1 & c_1 & u_1 & v_1 & w_1 \\ a_2 &
    b_2 & c_2 & u_2 & v_2 & w_2 \\ a_3 & b_3 & c_3 & u_3 & v_3 & w_3
   \\ a_4 & b_4 & c_4 & u_4 & v_4 & w_4 \\ a_5 & b_5 & c_5 & u_5 &
   v_5 & w_5 \\
\end{array} \right].
\end{equation}
It is a standard result that $\mbox{Res}(F,G,H)=0$ whenever $F=0$, $G=0$ and $H=0$ have a common nonzero solution \cite{Sturmfels}.  
We make fruitful use of this fact to analyze our MSCO equations in the next subsection.  Note that although system \eqref{msco1}--\eqref{msco3} involves three polynomials in two variables, we will employ a homogenization trick to rewrite the system in three variables and apply the above result.

{\bf Photon Orbits and MBOs: polynomials of one variable.}  Let $f(\xi)=a_{0}\xi^2 + a_{1}\xi + a_{2}$ and $g(\xi)=b_{0}\xi^2 + b_{1}\xi + b_{2}$ be two polynomials of the single variable $\xi$. The resultant of $f$ and $g$ is given by
\begin{equation}\label{eq:fg}
\mbox{Res}(f,g) = \det \left[ \begin{array}{cccc} a_0 & 0 & b_0 & 0
    \\ a_1 & a_0 & b_1 & b_0 \\ a_2 & a_1 & b_2 & b_1 \\ 0 & a_2 & 0 &
    b_2 \\
 \end{array}
 \right] .
\end{equation}
As above, it is also known that $f$ and $g$ have a common factor if and only if Res$(f,g)=0$.  See \cite{Cox}, for example.

We note that photon orbit conditions \eqref{eq:fullphoton} involve two polynomials in two variables, $E$ and $L$.  Defining a new variable in terms of the impact parameter (a ratio of $E$ and $L$), the problem reduces to that of finding common roots for two univariate polynomials.  Additionally, the marginally bound orbit conditions \eqref{eq:fullMBO} already appear as polynomial equations in $L$ alone.  In both instances, the above, simpler resultant treatment applies.

\subsection{Homogeneous orbit conditions}

{\bf MSCOs.}  To employ the results of the previous subsections, we must homogenize system
\eqref{msco1}--\eqref{msco3}.  For ease of notation we
make the identification $E=x$ and $L=y$ and set $\epsilon = -1$.  Then in each of
\eqref{msco1}, \eqref{msco2} and \eqref{msco3}, we replace $x \mapsto
x/z$, $y\mapsto y/z$ and multiply by $z^2$ to obtain the polynomials
$F,G$ and $H$ appearing in \eqref{eq:FGH}.  Their nonzero coefficients
are given below, as are the Jacobian derivative coefficients, for the
record:
\begin{multline}
a_0 = C, \, a_1 = -B, \, a_2 = -A, \, a_5 = - \rho^2, 
\\
b_0 = -\rho C^\prime + 2C, \, b_1 = \rho
B^\prime - 2B, \, b_2 =  \rho A^\prime - 2A,
\\
c_0 = -\rho^2 C^{\prime\prime} + 4\rho C^\prime - 6C,
\, c_1 =  \rho^2 B^{\prime\prime} - 4\rho B^\prime + 6B,
\\
c_2 = \rho^2 A^{\prime\prime} - 4\rho A^\prime +
6A, \, u_3  = 8a_5 (b_0 c_1-b_1 c_0),
\\
u_4   = 8a_5 (b_0 c_2-b_2 c_0), \, v_3 = u_4, \, v_4 = 8a_5 (b_1 c_2-b_2 c_1),
\\
w_0 = \frac{u_3}{2}, \quad w_1 = 8a_5 (b_0 c_2-b_2 c_0) , \quad w_2 = \frac{v_4}{2}.
\end{multline}
Calculating the resultant now becomes a straightforward task, yielding
\begin{eqnarray}\label{eq:ResultantFGH}
\!\!\!\mbox{Res}(F,G,H) 
 \!&=& \!a_5 \det \!\!\left[
 \begin{array}{cc} u_3 & v_3 \\
 u_4 & v_4
\end{array} \right]
 \! \det \!\!\left[ \begin{array}{ccc} b_0 & c_0 & w_0 \\ b_1 & c_1 & w_1
     \\ b_2 & c_2 & w_2
 \end{array} \right]\!.
\end{eqnarray}
After rewriting this in terms of the coefficients $a_0 , \ldots , c_5$ and observing that the two determinants appearing in \eqref{eq:ResultantFGH} are equal up to a constant multiple of $a_5$, the MSCO condition may be reduced to 
\begin{widetext}
\begin{eqnarray}\label{eq:BigOne}
0 &=& 
 \rho^4 \Big{\{} \left[ (
  \rho B^\prime - 2B)( \rho^2 C^{\prime\prime} - 4\rho C^\prime + 6C)
  - ( \rho C^\prime - 2C)( \rho^2 B^{\prime\prime} - 4\rho B^\prime +
  6B) \right]  \nonumber \\
   & & \left. \quad \qquad \cdot
\left[ ( \rho B^\prime - 2B)( \rho^2 A^{\prime\prime} - 4\rho A^\prime
  + 6A) - ( \rho A^\prime - 2A)( \rho^2 B^{\prime\prime} - 4\rho
  B^\prime + 6B) \right] \right.  \\ 
  & &  - \left[ (
  \rho A^\prime - 2A)( \rho^2 C^{\prime\prime} - 4\rho C^\prime + 6C)
  - ( \rho C^\prime - 2C)( \rho^2 A^{\prime\prime} - 4\rho A^\prime +
  6A) \right]^2 \Big{\}} . \nonumber
\end{eqnarray}
\end{widetext}
This equation gives a necessary and sufficient mathematical condition for existence of solutions to the algebraic system appearing in 
\eqref{eq:fullsystem}.  The open conditions $E^2>0$ and $L^2>0$ may then be imposed to determine the physicality of these candidate solutions.

We have assumed throughout the article that $\rho \neq 0$, so condition \eqref{eq:BigOne}
can be written more succinctly as
\begin{equation}\label{necessaryCondition}
\left[\mathcal{O}_{BC}-\mathcal{O}_{CB}\right]\left[\mathcal{O}_{BA}-\mathcal{O}_{AB}\right]-\left[\mathcal{O}_{AC}-\mathcal{O}_{CA}\right]^2=0
\end{equation}
where we define the operator $\mathcal{O}$ on two functions $F(\rho)$ and $G(\rho)$ by
\begin{equation}\label{OperatorMSCO}
\mathcal{O}_{FG} \equiv \left(\rho\frac{dF}{d\rho}-2F\right)\left(\rho^2\frac{d^2G}{d\rho^2}-4\rho\frac{dG}{d\rho}+6G\right).
\end{equation}
Observe that \eqref{necessaryCondition} reduces immediately to \eqref{determinantStaticCase} for static, axisymmetric spacetimes as a particular case ($\omega=0$, equivalently $a_1=b_1=c_1=0$).

\medskip


{\bf Photon Orbits and MBOs.}  As before, we may also derive a condition for the location of the photon orbit.  Observe that in the case $\epsilon =0$, the equations in \eqref{eq:fullphoton} form a homogeneous system of polynomial equations in $E=x$ and $L=y$.  Instead of using a homogenization procedure as above, we simplify this system to employ the single-variable resultant formulation given in the previous subsection.  We do this as follows: define an \emph{impact parameter} $b$ to be given by the ratio $b=|L/E|$.  Assuming $b \neq 0$, define $\xi \equiv x/y$. Then \eqref{eq:fullphoton} may be written as
\begin{eqnarray}\label{eq:photonpoly}
F(x,y) &:=& y^2(C \frac{x^2}{y^2} -BE \frac{x}{y} -A)  = 0, \nonumber \\
G(x,y) &:=& y^2 \Big{(} -\left[ \rho C^\prime - 2C \right] \frac{x^2}{y^2} + \left[ \rho
   B^\prime - 2B(\rho) \right] \frac{x}{y} \nonumber \\
   && \qquad + \left[ \rho A^\prime - 2A \right] \Big{)}= 0.
\end{eqnarray}
Thus, the polynomials $F(x,y)$, $G(x,y)$ in  \eqref{eq:photonpoly} can be rewritten as $F(x,y)= y^2f(\xi)$ and $G(x,y)=y^2g(\xi)$ with $f(\xi)=a_{0}\xi^2 + a_{1}\xi + a_{2}$ and
$g(\xi)=b_{0}\xi^2 + b_{1}\xi + b_{2}$. The resultant is then calculated via \eqref{eq:fg}.  After substitution of the coefficients in terms of the metric functions, the photon orbit condition reads
\begin{eqnarray}\label{PhotonOrbitStationaryCase}
(B A^\prime - B^\prime A)(B C^\prime - B^\prime C) - (A C^\prime - A^\prime C)  ^2 = 0
\end{eqnarray}
This last condition is precisely of the form appearing in \eqref{necessaryCondition} 
for $\mathcal{O}_{FG} \equiv FG^\prime$.  It is readily checked that in the static case ($B \equiv 0$) this expression reduces to the simple condition in \eqref{StaticPhotonOrbit}.

Analogously, we calculate the resultant of \eqref{eq:fullMBO} for marginally bound orbits in the stationary case as 
\begin{multline}\label{MBOStationaryCase}
\left[ ( \rho^2-C ) B^\prime - B (2\rho - C^\prime) \right]\cdot \left[ A^\prime B - A B^\prime \right] \\
- \left[ (\rho^2 - C) A^\prime - A (2\rho - C^\prime) \right]^2 = 0.
\end{multline}
This equation may also be rewritten in a form similar to \eqref{necessaryCondition}, namely
\begin{multline}\label{necessaryConditionMBO}
\left[\mathcal{O}_{P(C)B}-\mathcal{O}_{BP(C)}\right]\left[\mathcal{O}_{BA} -\mathcal{O}_{AB}\right] \\
-\left[\mathcal{O}_{P(C)A}-\mathcal{O}_{AP(C)}\right]^2=0,
\end{multline}
for $\mathcal{O}_{FG} \equiv FG^\prime$ and $P(F) \equiv \rho^2-F$.

\medskip

We are now in a position to apply resultant conditions \eqref{necessaryCondition} and
\eqref{PhotonOrbitStationaryCase} to obtain the location for the MSCOs,
 photon orbits, and MBOs for \emph{stationary}, axisymmetric spacetimes of
interest.

\subsection{ Kerr Spacetime}
The Kerr metric in Boyer-Lindquist coordinates reads
\begin{multline}
ds^2=\left( 1-\frac{2mr}{r^2 + a^2 \cos^2\theta}\right)^{-1} \Bigg[
  (r^2 -2mr + a^2)\sin^{2}\theta d\phi^2 \\ + \left(r^2 -2mr +
  a^2\cos^2\theta \right) \left(d\theta^2 + \frac{dr^2}{r^2-2mr +
    a^2}\right) \Bigg] \\ - \left(1-\frac{2mr}{r^2 +
  a^2\cos^2\theta}\right)\left(dt + \frac{2mr\sin^2\theta
  d\phi}{r^2-2mr+a\cos^2\theta }\right)^2.
\end{multline} 
Here, $m$ represents the total mass and $a=J/m$, where $J$ is the total angular momentum of the spacetime.  The black hole, extremal and hyperextremal cases are considered when
$m>a$, $m=a$ and $m<a$, respectively.  We transform to Weyl
coordinates via
\begin{equation}
\rho=\sqrt{r^2-2mr +a^2}\sin\theta, \qquad z=(r-m)\cos\theta.
\end{equation}
Using these two expressions, it is a straightforward computation to
find that at the equatorial plane $z=0$,
\begin{eqnarray}
 A(\rho)&=& \frac{-m+ \sqrt{\rho^2+m^2-a^2}}{ m+\sqrt{\rho^2+m^2-a^2}} \nonumber \\
 B(\rho) &=& -\frac{4ma}{m+\sqrt{\rho^2+m^2-a^2}} \\
 C(\rho) &=& \frac{-4m^2 a^2 + \rho^2 (m + \sqrt{\rho^2+m^2-a^2})}{\rho^2
   -a^2}. \nonumber
\end{eqnarray}
Substituting these expressions in the stationary MSCO condition
\eqref{necessaryCondition}, we obtain
\begin{equation}\label{eq:Fequation}
\frac{16m^2\rho^8
  F(\rho)}{(\rho^2+m^2-a^2)(m+\sqrt{\rho^2+m^2-a^2})^6}=0,
\end{equation}
where $F(\rho)$ is given by
\begin{eqnarray}
F(\rho) &=& \sqrt{\rho^2+ m^2-a^2}\left[64 a^4 m - 192 a^2 m^3 + 128 m^5  \right. \nonumber \\
&& \left. + (- 40 a^2 m + 32 m^3 ) \rho^2 - 6 m \rho^4 \right] + \rho^6 \nonumber \\
&& + (- 9 a^2 - 6 m^2 )\rho^4 + (24 a^4 -120 a^2 m^2 + 96 m^4 ) \rho^2   \nonumber \\
&& \qquad -16 a^6 + 144 a^4 m^2 - 256 a^2 m^4 + 128 m^6.
\end{eqnarray}
We claim that $F(\rho)=0$ encodes the MSCO condition for the Kerr spacetime.  Note that zeros of the denominator of \eqref{eq:Fequation} are located at $\rho^2 = a^2-m^2$ and $\rho^2 = m^2$ (corresponding to $r=m$ and $r=0,2m$, respectively), discussed below.

At the equatorial plane, $\rho=\sqrt{r^2-2mr+a^2}$, whereby $F(\rho)=0$ is equivalent to the equation
\begin{eqnarray}\label{LongOne}
&& |r-m| \left\{18a^4m^2-160a^2m^3+128m^5+104a^2m^2 \right.\nonumber \\
&&  \quad + (104a^2m^2 -64m^4)r + (8m^3- 52a^2m)r^2 \nonumber \\
&&  \quad \left. + 24m^2r^3 -6mr^4 \right\} + 18a^4m^2-160a^2m^4+128m^6 \nonumber \\
 && \quad  +
  (-18a^4m+264a^2m^3 -192m^5)r \nonumber \\ 
  && \quad + (9a^4-156a^2m^2
  +72m^4)r^4 + (24a^2m+16m^3)r^3 \nonumber \\ 
  && \quad+ 6(m^2-a^2)r^4-6mr^5 + r^6 =0.
\end{eqnarray}
Since the horizons are located at $r_{\mathcal{H^{\pm}}}=m \pm
\sqrt{m^2-a^2}$, we consider the region outside the outer horizon
$r>r_{\mathcal{H}^{+}}\geq m$, reducing \eqref{LongOne} to
\begin{equation}\label{eq:ourKerrMSCO}
r^2\left(9 a^4 - 28 a^2 m r - ( 6 a^2 - 36 m^2 ) r^2 - 12 m r^3 + r^4\right)=0.
\end{equation}
On the other hand, the equation for the MSCO in the equatorial plane for the Kerr
spacetime is known to be $-3a^2 + r^2 -6mr \mp 8 \sqrt{mr}=0$ (e.g., \cite{Mei02,PugQueRuf11b, Cha98}), which can be rearranged as the quartic in $r$
\begin{equation}\label{MSCOkerrLiteratureExpanded}
r^4 - 12 m r^3 + (- 6 a^2 + 36 m^2 ) r^2 - 28 a^2 m r + 9 a^4 = 0.
\end{equation}
Disregarding the $r=0$ solution, which corresponds to the location of
the singularity, \eqref{eq:ourKerrMSCO} visibly coincides with \eqref{MSCOkerrLiteratureExpanded}.

It is interesting to note that considering the case $r<m$ leads to a
reduction of \eqref{LongOne} to the polynomial equation
\begin{eqnarray}
&&(2m-r)^2\left[9a^4-80a^2m^2 +64m^4 +(52a^2m-32m^3)r \right. \nonumber \\
&&\qquad \qquad \qquad  \left. -6(a^2+ 2m^2)r^2 + 4mr^3 + r^4\right]=0.
\end{eqnarray}
The root $r=2m$ is automatically discarded since by assumption $r<m$,
leaving other candidates for roots encoded in the second factor.  It
should be mentioned that inside the horizon the spacetime character of
$r$ and $t$ are interchanged so that $r$-circular orbits correspond to
closed timelike geodesics.  Further analysis of these roots will be
left for future work.

For completeness, we verify that condition
\eqref{PhotonOrbitStationaryCase} for the location of the photon orbit
yields
\begin{align}\label{PhotonOrbitKerrInRhoCoords}
&0 = 4 a^2 m \rho^2 - 4 m^3 \rho^2 - 4 m \rho^4   \\
&+ \sqrt{\rho^2
   +m^2-a^2}(a^4 - a^2 m^2 - 2 a^2 \rho^2 + 5 m^2 \rho^2 + \rho^4). \nonumber
\end{align}
Again, at the equatorial plane $\rho=\sqrt{r^2-2mr+a^2}$, equation \eqref{PhotonOrbitKerrInRhoCoords} is rewritten as
\begin{equation}\label{photonOrbitInKerrTranslationOfOurResults}
(r-m)(r-2m)(- 4 a^2 m + 9 m^2 r - 6 m r^2 + r^3 )=0.
\end{equation}
Restricting to the region $r>r_{\mathcal{H}^{+}}$, we discard
the solution $r=m$ and find the location of the photon orbit in Kerr
spacetime encoded by the last term in the product.  The root $r=2m$ is
disregarded since it corresponds to the ergoregion determined by $r_{\pm}=m \pm \sqrt{m^2 -a^2 \cos^2 \theta}$
intersecting the equatorial plane $\theta=\pi/2$ at $r=m$.  See
\cite{PugQueRuf11b}.  This factor compares directly with the equation
found using the effective potential method.  The location
of the photon orbit in the equatorial plane of the Kerr spacetime
appearing in \cite{BarPreTeu72} is
\begin{equation}\label{photonOrbitKerrLiterature}
2 a \sqrt{m}-3m \sqrt{r} \mp r^{3/2}=0,
\end{equation}
for instance.  Under the above restrictions on $r$, this equation is equivalent to
$- 4 a^2 m + 9 m^2 r - 6 m r^2 + r^3=0$,
which directly compares with \eqref{photonOrbitInKerrTranslationOfOurResults}.

As in the previous examples, the marginally bound orbit condition is found via \eqref{MBOStationaryCase} in Weyl coordinates as
\begin{multline}
\sqrt{\rho^2+m^2-a^2}\left[ -8a^2m^3+8m^5-4m^3\rho^2 \right] \\
\quad + \left( 4a^4m^2-12a^2m^4+8m^6-4a^2m^2\rho^2+m^2\rho^4\right)=0. 
\end{multline}
As anticipated, transforming to Boyer-Lindquist coordinates recovers the classical Kerr MBO found in  \cite{Raine09}.

\subsection{Non-accelerating Pleba\'{n}ski-Demia\'{n}ski spacetime }
\label{PlebanskiDemianski}
A more general class of spacetimes that contains the Kerr metric as
a limiting case is the Pleba\'{n}ski-Demia\'{n}ski family.  Having already been cast in the Lewis-Papapetrou form \eqref{Weyl-Lewis-Papapetrou}, these spacetimes form the next natural candidates for the algorithmic method being put forward here; a complete discussion of the Pleba\'{n}ski-Demia\'{n}ski spacetime may be found in \cite{GriPod06}, for instance.  In the conventions and notation of this article,
 the metric is given locally by
 \begin{widetext}
\begin{eqnarray}\label{MetricCoefficientsPlebanski}
e^{2U} & = &  \frac{(r-m)^2 + a^2\cos^2\theta +q^2+g^2-m^2-l^2}{r^2+(l+a\cos\theta)^2}, \nonumber \\
   \omega & = & (1-\cos\theta) \frac{\left(a(1+\cos\theta)+2l\right)(2mr-q^2-g^2)+2l\left((a+l)(l+a\cos\theta)-r^2\right)}{(r-m)^2   +a^2\cos\theta + q^2 +g^2 -m^2 -l^2}, \nonumber \\ 
e^{-2U}e^{2\gamma} & = & \frac{(r-m)^2 + (a^2 + q^2 +g^2 -m^2 -l^2)\cos^2\theta}{r^2 + (l +  a\cos\theta)^2}. 
\end{eqnarray}
In the equatorial plane, this yields
\begin{eqnarray}
A &=& \frac{(r-m)^2 +q^2+g^2-m^2-l^2}{r^2+l^2} \nonumber \\
B &=& -\frac{2\left( (a+2l)(2mr-q^2-g^2)+2l \left(l(a+l)-r^2 \right) \right) }{r^2+l^2} \\
C &=& -\frac{\left[ (a+2l)(2mr-q^2-g^2)+2l(l(a+l)-r^2) \right]^2}{(r^2+l^2)((r-m)^2 +q^2+g^2-m^2-l^2)}+\frac{(r^2+l^2)\left((r-m)^2-z_{1}^2 \right)}{(r-m)^2 +q^2+g^2-m^2-l^2}. \nonumber 
\end{eqnarray}
\end{widetext}
In full generality, this family of spacetimes contains six parameters: $m$, $a$, $q$, $g$, $l$, and $\alpha$. The first four parameters are related to mass, angular momentum, electric and magnetic charges respectively, while $l$ is the NUT parameter and $\alpha$ represents acceleration of sources. For simplicity, we restrict our attention to the non-accelerating case $\alpha=0$, noting that a similar analysis applies for $\alpha \neq 0$.

Here, Boyer-Lindquist type coordinates $(r,\theta)$ are related to cannonical Weyl coordinates $(\rho,z)$ via
\begin{eqnarray} \label{changeCoordsPlebanski}
r & = & m + \frac{1}{2} \left(\sqrt{\rho^2 + (z-z_{1})^2}+ \sqrt{\rho^2 + (z-z_{2})^2}\right), \nonumber \\
\cos \theta & = & \frac{1}{2z_{1}}\left(\sqrt{\rho^2 + (z-z_{2})^2}- \sqrt{\rho^2 + (z-z_{1})^2}\right),
\end{eqnarray}
with $z_{1}=-z_{2}\equiv \sqrt{m^2 +l^2 -(a^2 + q^2 + g^2)}$.  Note also that the location of the equatorial plane $\theta$ is still determined by  $z=0$.

To compute the MSCO condition \eqref{necessaryCondition} in this spacetime, we first rewrite expression \eqref{OperatorMSCO} in terms of $\frac{d}{dr}$ as
\begin{eqnarray}\label{OperatorInBoyerLindquist}
\mathcal{O}_{FG} &=&
\left(\rho(r)\frac{dr}{d\rho}\frac{dF}{dr}-2F\right)\left(\rho^2(r)\left(\frac{dr}{d\rho}\right)^2
\frac{d^2G}{dr^2} \right. \nonumber \\
&& \qquad \left. + \left( \rho^2(r)
-4\rho(r)\frac{dr}{d\rho}\right)\frac{dG}{dr} + 6 G \right).
\end{eqnarray}
A direct computation using \eqref{changeCoordsPlebanski} and \eqref{OperatorInBoyerLindquist} (with $z=0$) 
renders the resultant as
\begin{equation}\label{eq:PlebMSCOfull}
\frac{-16(a^2+q^2+g^2-l^2-2mr-r^2)^4 F_{PD}(r,m,a,q,g,l)}{(r-m)^6(l^2+r^2)},
\end{equation}
where $F_{PD}(r,m,a,q,g,l)$ is a degree 12 polynomial in $r$ whose
coefficients depend only on $m$, $a$, $q$, $g$, and $l$.  The complete polynomial is provided in Appendix B.  Notice that the factor $\rho^2
=a^2+q^2+g^2-l^2-2mr-r^2$ in \eqref{eq:PlebMSCOfull} corresponds to the
location of the horizon.  Consequently, the MSCO condition for the
Plebanski-Demianski spacetime simply reads 
\begin{equation}\label{eq:PlebMSCO}
F_{PD}(r,m,a,q,g,l)=0,
\end{equation}
where $F_{PD}$ is given by \eqref{MSCOPlebanski}.

For the photon orbit analysis, notice that after rewriting the derivatives in \eqref{PhotonOrbitStationaryCase} with respect to $\rho$, the term
$(\frac{dr}{d\rho})^2$ appears as a common factor.  Assuming $\frac{dr}{d\rho} \neq 0$, the photon orbit condition reads formally identically to \eqref{PhotonOrbitStationaryCase} replacing the
derivatives respect to $\rho$ by derivatives respect to $r$.  A
straightforward computation then yields an equation for
the locus of the photon orbit, namely 
\begin{multline}\label{eq:PlebPhotonOrbit}
  r^6 -6m r^5 + ( 4 g^2 - 6 l^2 + 9 m^2 + 4 q^2)r^4 \\
  -4 m (a^2 + 3 g^2 - 5 l^2 + 3 q^2)r^3  + (4 g^4 + 9 l^4 - 6 l^2 m^2 - 12 l^2 q^2  \\
   + 4 q^4 + 4 a^2 (g^2 - 2 l^2  + q^2) + g^2 (-12 l^2 + 8 q^2))r^2 \\ 
  + 2 l^2 m (2 a^2 + 2 g^2 - 3 l^2 + 2 q^2)r + m^2l^4 =0.
\end{multline}

\medskip

We mention two significant subcases of \eqref{eq:PlebMSCO} and \eqref{eq:PlebPhotonOrbit}.  Setting the magnetic charge $g=0$ gives rise to a pair of simplified MSCO conditions for test particles and photon orbits, respectively.  The test particle MSCO is a degree 12 polynomial, written explicitly in Appendix B; the photon orbit above also reduces by setting $g=0$.  
A straightforward algebraic manipulation of the two polynomials confirms these to be the MSCO conditions of the Kerr-Newman-NUT spacetime reported in \cite{Pra15}.

Second, setting the NUT parameter $l=0$ in \eqref{eq:PlebMSCO} reduces the MSCO condition to 
\begin{multline}\label{MagneticBH}
m^2r^6 -12m^3r^5 + 6m^2(-a^2 + 3(g^2 +2m^2 +q^2))r^4 \\
-4m(a^2(-2g^2+7m^2-2q^2)  +(g^2+q^2)(2g^2 + 27m^2 + 2q^2))r^3 \\
+ 3m(3a^4 +30a^2(g^2 + q^2)+43(g^2 + q^2)^2)r^2 \\ 
-24m((g^2 + q^2)(a^4 + 4a^2(g^2 + q^2)+ 3(g^2 + q^2)^2))r \\
+16(g^2+q^2)^2(a^2+g^2+q^2)^2=0.
\end{multline}

To the knowledge of the authors, neither the MSCO condition given in \eqref{MSCOPlebanski} for the non-accelerating Pleba\'{n}ski-Demia\'{n}ski spacetime nor the MSCO equation in the simpler $l=0$ case in \eqref{MagneticBH} have been reported explicitly in the literature on hairy black holes.

Finally, a marginally bound orbit equation may also be found in full generality for this spacetime.  We use the same techniques as in the previous examples and suppress details of the calculation which yields this new MBO condition as a degree 8 polynomial in $r$; see Appendix B.

\subsection{Cylindrically symmetric spacetimes}

The algorithmic method put forward in this article can be succinctly
applied to stationary \emph{non-spherically symmetric} spacetimes
which can nonetheless be expressed in the Weyl-Papapetrou form.  We briefly demonstrate this, by computing MSCOs and photon orbits for solutions in the class of \emph{stationary, cylindrically symmetric} solutions. These solutions posses
a timelike Killing vector $\partial_{t}$  and two
spacelike commuting Killing vectors  $\partial_{\phi}$ and $\partial_{z}$.
  These solutions can be used to describe the exterior region 
of a rotating cylindrical source.  See \cite{SKMHH, McCal98, CarSen99, McCalSan98} for 
comprehensive discussion of such metrics.

 \emph{Kasner spacetimes.}  A vacuum solution in the static, cylindrically-symmetric class is given by 
\begin{equation}
ds^2=\rho^{-2a}\left( \rho^{2a^2}(d\rho^2 + dz^2) + \rho^2
d\phi^2\right)-\rho^{2a}dt^2.
\end{equation}
These are Kasner solutions which are flat for $a=0,1$ and Petrov type
D for $a=2, 1/2,-1$.  See \cite{SKMHH, Bonnor79}.  For this spacetime,  \eqref{necessaryCondition} becomes
 $a^2(1-3a+2a^2)^2\rho^4=0$.
 Thus, for $a \neq 0,1, 1/2$, the only possible candidate for a MSCO on
the equatorial plane is $\rho=0$, whereas for $a=0,1,1/2$, condition
\eqref{necessaryCondition} is trivially satisfied for any $\rho$.
Using \eqref{PhotonOrbitStationaryCase} we also easily find the photon
orbit equation 
 $(1-2a)^2 \rho^4=0$.
 Consequently, for $a \neq 1/2$, the photon orbit is located at
$\rho=0$, while for $a=1/2$ the condition
\eqref{PhotonOrbitStationaryCase} is satisfied for any $\rho$, as expected.

\emph{Islam-Van den Bergh-Wils spacetime.}  The Islam-Van den Bergh-Wils spacetime is a stationary, 
cylindrically-symmetric solution to the Einstein-Maxwell field
equation described by line element and vector potential
\begin{eqnarray}
ds^2 & = &\rho^{-4/9}\exp(-a^2\rho^{2/3})(d\rho^2 +
dz^2) \nonumber \\
&& \qquad +\rho^{4/3}d\phi^2 -\rho^{2/3}(dt + a \rho^{2/3}d\phi)^2,
\nonumber \\ 
A_{\alpha}dx^{\alpha} & =&
a\rho^{2/3}/\sqrt{2}+a^2\rho^{4/3}d\phi/\sqrt{8},
\end{eqnarray}
for constant $a$.  Straight forward application of conditions
\eqref{necessaryCondition} and \eqref{PhotonOrbitStationaryCase} shows
that the MSCO is located at 
 $\rho^2 - 36a^2 = 0$, 
 while the photon orbit condition is given by
 $\rho^2 - 4 a^2 (2 + \rho^{8/3} ) + 4 a^4 \rho^{2/3} (3 + \rho^{8/3})=0$. 
 Defining $x=\rho^{2/3}$ one obtains the quintic equation $4a^4x^5-4a^2x^4+x^3+12a^4x-8a^2=0$.  Several techniques are readily available for deducing that there is at least one positive root for $x$, and consequently $\rho$ (e.g., Descartes' rule of signs, Sturm's Theorem, numerical methods, etc.).

\section{Conclusions}
Our goal is to present a geometric alternative to the effective potential method for computing ISCOs/OSCOs/MBOs as well as to provide an algorithm which can be easily implemented to compute circular and parabolic orbits for a large class of metrics.  We derive necessary and sufficient mathematical conditions for the
existence of candidates for marginally stable circular orbits (MSCO) of test
particles in the gravitational field of stationary, axisymmetric
spacetimes possessing a reflection symmetry respect to the
equatorial plane.  Using resultants, the MSCO conditions assume an elegant form as  simple algebraic equations involving solely geometric data (i.e., metric coefficients) and the open physical conditions on energy and angular momentum provide a simple tool with which to check physicality of solutions to the algebraic system.  The method also applies for circular photon orbits and marginally bound orbits.

 In contrast to the effective potential method, which is typically applied on a case-by-case basis, our formulas can be systematically applied
  to any metric written in the appropriate (Weyl-Lewis-Papapetrou)
  form.  We demonstrate their simplicity by recovering determinantal formulas for static, spherically symmetric spacetimes as well as calculating
  explicit examples ranging from recovery of MSCOs, photon orbits, and MBOs for Schwarzschild
  and Kerr spacetimes to more exotic and new candidates such as
   spacetimes described by the q-metric and the non-accelerating 
 Pleba\'nski-Demia\'nski family of metrics.
 
  It is worth emphasizing that the approach put forward in
  this article gives an algorithmic procedure for determining
  MSCOs, photon orbits and MBOs for a given spacetime.  Pseudocode is provided in Appendix C for performing such calculations efficiently.  The general conditions derived in this article can be used not only to determine the MSCOs for various spacetimes of physical
interest, but also provide geometric insight into the properties of such orbits.  In the static, axisymmetric setting, the MSCO condition is completely characterized by the norm of the timelike Killing field, for instance.  Furthermore, this approach also guarantees that \emph{all
candidates} for MSCOs are identified.

Further understanding of the reduction procedure could have interesting observational consequences in astrophysics, as matter configuration in accretion is typically modelled
using stationary, axisymmetric spacetimes; one might attempt to exhibit explicit
dependence of the geometric structure of the accretion disk to
physical parameters determining the gravitational field by using more sophisticated algebro-geometric tools.  It may be possible, for instance, to use general properties of Groebner Bases--of which the resultant is one representative--to characterize the algebraic system which determines existence of MSCOs in terms of its geometry.  This will be pursued in a future work.
  
%

\begin{acknowledgments}
The authors would like to thank Juan Valiente Kroon for helpful comments and suggestions.  EG gratefully acknowledges support from Consejo Nacional de Ciencia y Tecnonolog\'ia (CONACyT Scholarship 494039/218141).  SB also thanks the London Mathematical Society for their support (Grace Chisholm Young Fellowship No. GCY 13-14 02).
\end{acknowledgments}

\appendix

\section{}\label{AppendixB}
In this section, we argue that if a stationary, axially symmetric
spacetime possesses a discrete reflection symmetry with respect to the
equatorial plane $z=0$, then under the appropriate initial conditions,
the geodesic motion of a particle will be constrained to this plane.
Let $g_{ab}$ denote the components of the metric in
Lewis-Papapetrou form, with respect to the coordinate basis $\{
\partial_{t},\partial_{\rho},\partial_{z},\partial_{\phi}\}$.  Let
$x^{a}(\tau)$ denote a timelike geodesic parametrised with respect to
an affine parameter $\tau$.  Assume the mass of the particle is given
by $m$ so that the momentum is given by $p_{a}=m v_{a}$, for $v^{a}={d
  x^{a}}/{d\tau}$.

Observe that the geodesic equation $v^{a}\nabla_{a}v^{b}=0$ can be
written in terms of the momentum $p_{a}$ as
$\frac{d}{d\tau}p_{a}=\frac{1}{2}(\partial_{a}g_{bc})p^{b}p^{c}$.
  Therefore, the momentum $p_{z}$ transversal to the plane $z=z_0$
satisfies
$\frac{d}{d\tau}p_{z}=\frac{1}{2}\left (\frac{\partial g_{bc}
}{\partial z} \right )p^{b}p^{c}$.
  If the metric  possesses a discrete reflection symmetry
respect to the equatorial plane, then $g_{ab}(t,\rho,z,\phi) =g_{ab}(t,\rho,-z,\phi)$. Consequently
$\frac{\partial}{\partial z}g_{ab}=0$ at $z=0$. Hence $\frac{ d
  p_{z}}{d\tau}=0$ at the equatorial plane $z=0$ and
$p_{z}(\tau)=p_{z}(\tau_0)$. Thus, if the momentum transversal to the
equatorial plane $p_{z}$ vanishes initially, it will vanish for later
times.


\section{}\label{AppendixC}

\textbf{Marginally Stable Circular Orbit (MSCO):}
The computation described in Section \ref{PlebanskiDemianski} leads to the following
condition for the location of the equatorial MSCOs
 in the Pleba\'nski-Demia\'nski spacetime
\begin{multline}\label{MSCOPlebanski}
F_{PD}(r,m,a,q,g,l) \equiv a_{12}r^{12} +a_{11}r^{11} + a_{10}r^{10} \\
+ a_{9}r^{9} + a_{8}r^{8}+ a_{7}r^{7}+a_{6}r^{6}  +a_{5}r^5 \\
+ a_{4}r^{4}+ a_{3}r^3+ a_{2}r^2 + a_{1}r + a_{0}=0.
\end{multline}
The coefficients $a_{i}$ ($i = 0 \ldots12$) appearing above depend only
on the parameters $m$, $a$, $q$, $g$, and $l$, and are given explicitly below:
\begin{widetext}
\begingroup
\allowdisplaybreaks
\begin{eqnarray*}
 a_{12} & = & m^2 \\
a_{11} & = & -12m^3  \\
a_{10} & = & 6 m^2 (- a^2 + 3 g^2 - 5 l^2 + 6 m^2 + 3 q^2) \\
a_{9} & = & -4 m \bigl(2 g^4 + 8 l^4 - 47 l^2 m^2 - 8 l^2 q^2 + 27 m^2 q^2 + 2 \
q^4 + a^2 (-2 g^2 + 4 l^2 + 7 m^2 - 2 q^2)  \\ && + g^2 (-8 l^2 + 27 m^2 + 4 \
q^2)\bigr)q^2)\bigr) \\
a_{8} & = & 3 m^2 \bigl(3 a^4 + 43 g^4 + 149 l^4 - 16 l^2 m^2 - 158 l^2 q^2 + 43 \
q^4 + a^2 (30 g^2 - 62 l^2 + 30 q^2) \\ &&  + g^2 (-158 l^2 + 86 q^2)\bigr) \\
a_{7} & = & -24 m \Bigl(3 g^6 - 20 l^6 + 13 l^4 m^2 + 32 l^4 q^2 - 6 l^2 m^2 q^2 \
- 17 l^2 q^4 + 3 q^6 + a^4 \bigl(g^2 - 2 l^2 + q^2\bigr)  \\ && + g^4 \
\bigl(-17 l^2 + 9 q^2\bigr) + g^2 \bigl(32 l^4 + 9 q^4 - 2 l^2 (3 m^2 \
+ 17 q^2)\bigr)  \\ && + a^2 \bigl(4 g^4 + 18 l^4 + 4 q^4 + g^2 (-17 l^2 + 8 \
q^2) -  l^2 (4 m^2 + 17 q^2)\bigr)\Bigr) \\
a_{6} & = & 4 \biggl(4 g^8 + 64 l^8 - 145 l^6 m^2 + 22 l^4 m^4 - 128 l^6 q^2 + \
145 l^4 m^2 q^2 + 96 l^4 q^4 - 35 l^2 m^2 q^4 \\ && - 32 l^2 q^6 + 4 q^8 + \
16 g^6 \Bigl(-2 l^2 + q^2\Bigr) + a^4 \Bigl(4 g^4 + 16 l^4 + 4 q^4 + \
8 g^2 \bigl(-2 l^2 + q^2\bigr)  \\ && -  l^2 \bigl(9 m^2 + 16 \
q^2\bigr)\Bigr) + g^4 \Bigl(96 l^4 + 24 q^4 -  l^2 \bigl(35 m^2 + 96 \
q^2\bigr)\Bigr) \\ && + g^2 \Bigl(-128 l^6 + 16 q^6 + l^4 \bigl(145 m^2 + \
192 q^2\bigr) - 2 l^2 \bigl(35 m^2 q^2 + 48 q^4\bigr)\Bigr)  \\ && + a^2 \
\Bigl(8 g^6 - 64 l^6 + 8 q^6 + 24 g^4 \bigl(-2 l^2 + q^2\bigr) + l^4 \
\bigl(109 m^2 + 96 q^2\bigr) \\ && - 4 l^2 \bigl(13 m^2 q^2 + 12 q^4\bigr) \
+ 4 g^2 \bigl(24 l^4 + 6 q^4 -  l^2 (13 m^2 + 24 \
q^2)\bigr)\Bigr)\biggr) \\
a_{5} & = & 24 l^2 m \Bigl(2 g^6 - 20 l^6 + 13 l^4 m^2 + 28 l^4 q^2 - 7 l^2 m^2 \
q^2 - 13 l^2 q^4 + 2 q^6  + 2 a^4 \bigl(g^2 - 2 l^2 + q^2\bigr) \\ && + g^4 \
\bigl(-13 l^2 + 6 q^2\bigr) + a^2 \bigl(4 g^4 + 14 l^4 + 4 q^4 - 5 \
l^2 (m^2 + 3 q^2)  + g^2 (-15 l^2 + 8 q^2)\bigr)  \\ && + g^2 \bigl(28 l^4 + \
6 q^4 -  l^2 (7 m^2 + 26 q^2)\bigr)\Bigr) \\
a_{4} & = & 3 l^4 m^2 \bigl(10 a^4 + 34 g^4 + 149 l^4 - 16 l^2 m^2 - 140 l^2 q^2 \
+ 34 q^4 + 4 a^2 (7 g^2 - 11 l^2 + 7 q^2) \\ &&  + g^2 (-140 l^2 + 68 \
q^2)\bigr) \\
a_{3} & = & 4 l^4 m \Bigl(-2 g^6 + 8 l^6 - 47 l^4 m^2 - 16 l^4 q^2 + 20 l^2 m^2 \
q^2 + 10 l^2 q^4 - 2 q^6 + 2 g^4 \bigl(5 l^2 - 3 q^2\bigr) \\ && + 2 a^4 \
\bigl(g^2 - 2 l^2 + q^2\bigr) + 2 a^2 l^2 \bigl(g^2 - 2 l^2 + \
q^2\bigr) + g^2 \bigl(-16 l^4 - 6 q^4 + 20 l^2 (m^2 + q^2)\bigr)\Bigr) \\
a_{2} & = & 6 l^6 m^2 \bigl(2 a^4 - 2 g^4 + 3 a^2 l^2 - 5 l^4 + 6 l^2 m^2 + 7 l^2 \
q^2 - 2 q^4 + g^2 (7 l^2 - 4 q^2)\bigr)\\
a_{1} & = & -12 l^8 m^3 (a^2 + g^2 -  l^2 + q^2) \\
a_{0} & = & l^8 m^2 (a^2 + g^2 -  l^2 + q^2)^2.
\end{eqnarray*}
 \endgroup
 \end{widetext}
Observe that by setting the magnetic charge $g=0$, the condition $F_{PD}(r,m,a,q,g,l)=0$ reduces
to the MSCO condition for the Kerr-Newman-NUT spacetime.
After some algebra, equation $F_{PD}(r,m,a,q,g=0,l)=0$ determined by the reduced coefficients is found to be equivalent to Equation 81 appearing in \cite{Pra15}.

\textbf{Marginally bound orbit (MBO):}
The MBO condition for the Pleba\'nski-Demia\'nski spacetime is determined by the following polynomial equation
\begin{multline}
H_{PD}(r,m,a,q,g,l) \equiv  b_{8}r^{8}+ b_{7}r^{7}+b_{6}r^{6} +b_{5}r^5\\
+ b_{4}r^{4}+ b_{3}r^3+ b_{2}r^2 + b_{1}r + b_{0} =0, 
\end{multline}
where the coefficients are given explicitly by
\begin{widetext}
\begingroup
\allowdisplaybreaks
\begin{eqnarray*}
b_{8} & = & 4 m^2 \\
b_{7} & = & -32 m^3 \\
b_{6} & = & -2 a^2 m^2 + 8 g^2 m^2 - 12 l^2 m^2 + 16 m^4 + 8 m^2 q\\
b_{5} & = & 4 (2 a^2 g^2 m - 2 g^4 m - 4 a^2 l^2 m + 8 g^2 l^2 m - 8 l^4 m - 8 \
a^2 m^3 \\  && - 32 g^2 m^3 + 48 l^2 m^3 + 2 a^2 m q^2 - 4 g^2 m q^2 + 8 l^2 \
m q^2 - 32 m^3 q^2 - 2 m q^4) \\
b_{4} & = & 4 (a^4 m^2 + 16 a^2 g^2 m^2 + 24 g^4 m^2 - 34 a^2 l^2 m^2 - 80 g^2 \
l^2 m^2 + 70 l^4 m^2 + 16 a^2 m^2 q^2 \\&& + 48 g^2 m^2 q^2  - 80 l^2 m^2 \
q^2 + 24 m^2 q^4) \\
 b_{3} & = &  4 (-2 a^4 g^2 m - 10 a^2 g^4 m - 8 g^6 m + 4 a^4 l^2 m + 44 a^2 g^2 \
l^2 m + 44 g^4 l^2 m - 48 a^2 l^4 m \\&& - 80 g^2 l^4 m  + 48 l^6 m + 8 a^2 \
l^2 m^3 - 8 l^4 m^3 - 2 a^4 m q^2 - 20 a^2 g^2 m q^2 - 24 g^4 m q^2 + \\ && \
44 a^2 l^2 m q^2 + 88 g^2 l^2 m q^2 - 80 l^4 m q^2 - 10 a^2 m q^4 - \
24 g^2 m q^4 + 44 l^2 m q^4 - 8 m q^6) \\
b_{2} & = & 4 (a^4 g^4 + 2 a^2 g^6 + g^8 - 4 a^4 g^2 l^2 - 12 a^2 g^4 l^2 - 8 g^6 \
l^2 + 4 a^4 l^4 + 24 a^2 g^2 l^4 + 24 g^4 l^4 \\ && - 16 a^2 l^6  - 32 g^2 \
l^6 + 16 l^8 - 2 a^4 l^2 m^2 - 8 a^2 g^2 l^2 m^2 + 18 a^2 l^4 m^2 + 8 \
g^2 l^4 m^2 \\ && - 12 l^6 m^2 +  2 a^4 g^2 q^2 + 6 a^2 g^4 q^2  + 4 g^6 q^2 \
- 4 a^4 l^2 q^2 - 24 a^2 g^2 l^2 q^2 - 24 g^4 l^2 q^2 \\ && + 24 a^2 l^4 \
q^2  + 48 g^2 l^4 q^2 - 32 l^6 q^2 - 8 a^2 l^2 m^2 q^2 + 8 l^4 m^2 q^2 \
+ a^4 q^4 + 6 a^2 g^2 q^4 \\ && + 6 g^4 q^4  - 12 a^2 l^2 q^4 - 24 g^2 l^2 \
q^4 + 24 l^4 q^4 + 2 a^2 q^6 + 4 g^2 q^6 - 8 l^2 q^6 + q^8)  \\
b_{1} & = & 4 (2 a^4 g^2 l^2 m + 2 a^2 g^4 l^2 m - 4 a^4 l^4 m - 6 a^2 g^2 l^4 m \
- 2 g^4 l^4 m + 4 a^2 l^6 m \\ && +  8 g^2 l^6 m - 8 l^8 m + 2 a^4 l^2 m q^2 \
+ 4 a^2 g^2 l^2 m q^2 - 6 a^2 l^4 m q^2 - 4 g^2 l^4 m q^2 \\ && + 8 l^6 m \
q^2 + 2 a^2 l^2 m q^4 - 2 l^4 m q^4)\\
b_{0} & = & 4 (a^4 l^4 m^2 + 2 a^2 l^6 m^2 + l^8 m^2).
\end{eqnarray*}
 \endgroup
\end{widetext}

\section{}\label{AppendixD}

There are many ways to calculate resultants using computer algebra packages.  In this Appendix, we provide Mathematica pseudocode as one possible method for reproducing the computations of this paper.  Note that under the predefined ordering used in the algebraic geometry package of Mathematica, the first polynomial appearing as output of the Groebner Basis command is precisely the resultant defined in Section 4.

\begin{widetext}
{\small
\verb| (*MSCO*)|

\verb|Clear[\[ScriptCapitalA],\[ScriptCapitalB],\[ScriptCapitalC]] |

\medskip

\verb| (*Effective potential*) |

\verb|\[CapitalPhi][\[Rho]_] := (1/\[Rho]^2)*(-\[ScriptCapitalC][\[Rho]]*\\[DoubleStruckCapitalE]^2|

\verb|+\[ScriptCapitalB][\[Rho]]*\\[DoubleStruckCapitalE]*\[DoubleStruckCapitalL] |

\verb|+\[ScriptCapitalA][\[Rho]]*\\[DoubleStruckCapitalL]^2) |

\medskip

\verb| (*MSCO system*) |

\verb|Collect[Expand[\[Rho]^2*(\[Eps]-\[CapitalPhi][\[Rho]])],{\\[DoubleStruckCapitalE],\[DoubleStruckCapitalL]}]|

\verb|Collect[Expand[\[Rho]^3*\[CapitalPhi]'[\[Rho]]],{\\[DoubleStruckCapitalE],\[DoubleStruckCapitalL]}]|

\verb|Collect[Expand[\[Rho]^4*\[CapitalPhi]''[\[Rho]]],{\\[DoubleStruckCapitalE],\[DoubleStruckCapitalL]}]|

\medskip

\verb| (*This is viewed as a polynomial system in {\[DoubleStruckCapitalE],\[DoubleStruckCapitalL]}*)|

\verb| (*To simplify the notation let x = \[DoubleStruckCapitalE] and y = \[DoubleStruckCapitalL]. Finding MSCO|

\verb| conditions is  equivalent to finding conditions for the polynomials {f, g, h} to have common roots.*)| 

\verb|f[x_, y_] :=   Collect[Expand[\[Rho]^2*(\[Eps] - \[CapitalPhi][\[Rho]])], |

\verb|{\\[DoubleStruckCapitalE],\[DoubleStruckCapitalL]}]/.{\\[DoubleStruckCapitalE]->x,\[DoubleStruckCapitalL]->y}|

\verb|g[x_, y_] :=  Collect[Expand[\[Rho]^3*\[CapitalPhi]'[\[Rho]]], |

\verb|{\\[DoubleStruckCapitalE],\[DoubleStruckCapitalL]}]/.{\\[DoubleStruckCapitalE]->x,\[DoubleStruckCapitalL]->y}|

\verb|h[x_, y_] :=  Collect[Expand[\[Rho]^4*\[CapitalPhi]''[\[Rho]]], |
 
\verb|{\\[DoubleStruckCapitalE],\[DoubleStruckCapitalL]}]/.{\\[DoubleStruckCapitalE]->x,\[DoubleStruckCapitalL]->y}|

\verb|StationaryMSCO = Factor[GroebnerBasis[{f[x,y],g[x,y],h[x,y]},{x,y}][[1]]/(\[Eps]^2*\[Rho]^2)]|

\medskip

\verb| (*Photon Orbit*) |

\verb|Clear[\[ScriptCapitalA], \[ScriptCapitalB], \[ScriptCapitalC]] |

\medskip

\verb| (*Photon Orbit System*) |

\verb|f1photon[x_, y_] :=  Collect[Expand[\[Rho]^2*(\[CapitalPhi][\[Rho]])],|

\verb|{\\[DoubleStruckCapitalE],\[DoubleStruckCapitalL]}]/.{\\[DoubleStruckCapitalE]->x,\[DoubleStruckCapitalL]->y}|

\verb|f2photon[x_, y_] :=  Collect[Expand[\[Rho]^3*\[CapitalPhi]'[\[Rho]]],|

\verb|{\\[DoubleStruckCapitalE],\[DoubleStruckCapitalL]}]/.{\\[DoubleStruckCapitalE]->x,\[DoubleStruckCapitalL]->y}|

\medskip

\verb| (*Assume \[DoubleStruckCapitalL] nonzero, introduce impact parameter to obtain photon orbit*)|

\verb|F1photon[z_] := Expand[(1/y^2) *f1photon[z*y, y]], F2photon[z_] := Expand[(1/y^2) *f2photon[z*y, y]]|

\verb|PhotonOrbitStationary = Resultant[F1photon[z], F2photon[z], z];|

\medskip

\verb| (* Example: Extremal Kerr spacetime*)|

\verb|Clear[\[ScriptCapitalA], \[ScriptCapitalB], \[ScriptCapitalC]]|

\verb|\[ScriptCapitalA][\[Rho]_] := (\[Rho] - m)/(\[Rho] + m)|

\verb|\[ScriptCapitalB][\[Rho]_] := -4*m^2/(\[Rho] + m)|

\verb|\[ScriptCapitalC][\[Rho]_] := (\[Rho]^2*(\[Rho] + m)^2 - 4*m^4)/((\[Rho] + m) (\[Rho] - m))|


\verb|FullSimplify[StationaryMSCO]|

\verb|FullSimplify[PhotonOrbitStationary]|

\medskip

\verb| (*\[Rho]= 8m corresponds to r = 9 m which is the msco in extremal Kerr spacetime while \[Rho]= 3m|

\verb| corresponds to r = 4 m which is the location of the photon orbit in extremal Kerr*)|
}
\end{widetext}


\bibliography{MSCO_bib}

\end{document}